\documentclass[preprint,number,times,10xpt]{elsarticle}
\biboptions{sort&compress}
 
\usepackage[utf8]{inputenc}   
\usepackage[T1]{fontenc}      
\usepackage{xcolor}

\definecolor{darkred}{rgb}{0.65, 0, 0}
\definecolor{darkorange}{rgb}{0.8, 0.4, 0}

\usepackage{amsmath,amssymb,graphicx,bm}
\usepackage{physics}
\usepackage{slashed}
\usepackage{mathtools}
\usepackage{hyperref}
\usepackage{orcidlink}
\usepackage{tikz}
\usepackage{pgfplots}
\usepackage[normalem]{ulem}
\usepackage{subcaption}
 
\pgfplotsset{compat=1.18}
 
\newcommand{\Lag}{\mathcal{L}}

\renewcommand{\dd}{\mathrm{d}}

\newif\ifcalcsinbody
\calcsinbodytrue

\journal{Physics of the Dark Universe}
 
\begin{document}
 
\begin{frontmatter}
 
\title{Collider Probes of Dark Energy Microphysics}
 
\author[vandy]{Chaitanya Bashyam\,\orcidlink{0000-0002-7447-8651}}
\ead{chaitanya.bashyam@vanderbilt.edu}
 
\author[vandy]{Alfredo Gurrola\corref{cor1}\,\orcidlink{0000-0002-2793-4052}}
\ead{alfredo.gurrola@vanderbilt.edu}
\cortext[cor1]{Corresponding author.}
 
\affiliation[vandy]{organization={Department of Physics and Astronomy, Vanderbilt University},
            city={Nashville}, state={TN}, country={USA}}
 
\author[uandes]{Andrés Flórez\,\orcidlink{0000-0002-3222-0249}}
\ead{ca.florez@uniandes.edu.co}
 
\author[uandes]{Cristian Rodríguez\,\orcidlink{0000-0001-5166-2176}}
\ead{c.rodriguez45@uniandes.edu.co}
 
\affiliation[uandes]{organization={Department of Physics, Universidad de los Andes},
            city={Bogotá}, country={Colombia}}
 
\begin{abstract}
The physical origin of dark energy remains one of the most profound open questions in modern physics.
Although cosmological observations tightly constrain the equation of state parameter $w$, this information alone does not reveal the underlying microphysics, as many distinct theoretical models can reproduce the same expansion history. A key discriminator among these models is the sound speed of dark energy perturbations, yet this quantity remains largely unconstrained by current astrophysical observations. In this work, we propose a fundamentally new approach: using collider measurements of beyond-the-Standard-Model (BSM) mediator resonances as a probe of dark energy microphysics. We construct a unified effective-field-theory framework in which a dynamical dark energy scalar is coupled, through symmetry-motivated derivative interactions, to a pseudoscalar mediator in the 2HDM+$a$ model. These interactions naturally induce invisible decays and modify the propagation of the BSM mediator in a dark energy background, leading to measurable distortions of resonance properties at colliders such as the LHC. We show that the decay widths, branching ratios, and kinematic structure of the mediator resonance become sensitive to the propagation properties of dark energy fluctuations, in particular the sound speed.
As a result, collider observables provide a direct and complementary handle on dark energy microphysics, with the potential to distinguish between models that are otherwise indistinguishable through cosmology alone. Our results establish a new paradigm in which high-energy collider experiments can probe the physics of cosmic acceleration, revealing a connection between the smallest and largest scales in nature and opening a novel experimental pathway to uncover the fundamental origin of dark energy.
\end{abstract}
 
\end{frontmatter}

\section{Introduction}
The discovery that the expansion of the Universe is accelerating revealed the presence of a dominant component of energy density, dark energy, whose microscopic origin remains one of the deepest open questions in fundamental physics \cite{de1SupernovaSearchTeam:1998fmf,perlmutter1999measurements}. 
Despite decades of intense theoretical and observational effort, existing probes of dark energy rely almost entirely on cosmological measurements of the background expansion history or large-scale structure, leaving its underlying microphysics largely unconstrained \cite{weinberg2013observational,copeland2006dynamics,amendola2010dark}. 
In parallel, high-energy particle colliders such as the Large Hadron Collider (LHC) provide a powerful laboratory for exploring new degrees of freedom and interactions beyond the Standard Model (SM) \cite{atlas2012observation,cms2012observation}. 
Remarkably, these two frontiers of physics, cosmic acceleration and collider experiments, have traditionally been treated as essentially disconnected. 
In this work we explore the possibility that they may instead be directly linked: if dark energy arises from a dynamical scalar field that interacts with new particles accessible at collider energies, then measurements of new mediator resonances at colliders such as the LHC could provide a novel probe of the microphysics responsible for cosmic acceleration \cite{carroll1998quintessence,ferreira1998quintessence,burrage2018tests,brax2004detecting}.

Understanding the physical origin of cosmic acceleration remains one of the central challenges of modern cosmology. 
While the $\Lambda$CDM model successfully describes a wide range of observations, including the cosmic microwave background (CMB), large-scale structure, and Type Ia supernovae~\cite{de1SupernovaSearchTeam:1998fmf}, the nature of dark energy -- which dominates the energy density of the Universe at late times -- remains unknown. 
The cosmological constant interpretation of dark energy provides an excellent phenomenological fit to the data but raises profound theoretical puzzles, most notably the extreme discrepancy between the small observed vacuum energy and the particle-physics expectation~\cite{Avsajanishvili:2024obsconst_dynamde}.

This tension has motivated a large number of alternative scenarios ranging from modified gravity theories to dynamical scalar-field models and quantum-gravitational explanations of vacuum energy~\cite{de2Li:2012dt,de3Li:2011sd,Mortonson:2013zfa,de6Huterer:2017buf}. Many of these alternatives describe dark energy as a dynamical scalar field $\phi$ whose dynamics are governed by a nontrivial kinetic structure. 
A broad and widely used description is provided by k-essence models, which we treat throughout as an effective field theory (EFT) for a shift-symmetric scalar, valid below a cutoff energy scale $\Lambda$ \cite{Cheung:2007st,eft1gubitosi2013effective,Babichev:2018twg,Glavan:2025khe}.

In such theories, the Lagrangian function $P$ controls both the background evolution and the dynamics of perturbations. 
While the former is captured at leading order by the background equation-of-state parameter $w$, defined as the ratio of pressure to energy density for a given component ($w \equiv p/\rho$), vastly different microphysical models can share nearly identical background expansion histories~\cite{de7Vagnozzi:2021quy,de8Adil:2023ara,de10DiValentino:2020evt,de11Nojiri:2010wj,de12Nojiri:2006ri,Bamba:2012cp,DESI:2025zgx}. 
As a result, measurements of $w$ alone cannot uniquely determine the physical origin of cosmic acceleration, even though background data already disfavor part of the model space~\cite{Colgain:2025nzf}.

The dynamics of perturbations is governed by the sound speed of dark energy $c_s^2$, which controls the propagation of pressure perturbations and determines whether it clusters gravitationally on cosmological scales. 
In canonical scalar-field models one typically finds $c_s^2=1$, implying that dark energy remains nearly homogeneous. 
By contrast, a wide range of non-canonical theories -- including coupled dark energy models~\cite{cde1gumjudpai2005coupled,cde2gomez2020update,cde3barros2019kinetically,cde4xia2009constraint,cde5maccio2004coupled,Silva:2025hxw,Shah:2025ayl,Pan:2025qwy}, k-essence scenarios~\cite{nc1armendariz2001essentials,nc2malquarti2003new,nc3armendariz2005haloes,nc4ahn2009dark,nc5chimento2010dbi,nc6cai2016dark,nc7chattopadhyay2010interaction,nc8calcagni2006tachyon,nc9bagla2003cosmology,nc10copeland2005needed,nc11micheletti2010observational,nc12sheykhi2012tachyon,Kehayias:2019gir,Chiba:2009nh,Das:2006cm}, and EFT extensions of gravity~\cite{de5Frusciante:2019xia,eft1gubitosi2013effective,eft2linder2016effective,eft3bloomfield2013dark,eft4liang2023dark,eft5frusciante2014effective} -- naturally predict $c_s^2\ll1$ and, in some cases, superluminal propagation~\cite{Babichev:2007dw}.
Measuring or constraining this quantity therefore provides a direct window into the microphysics underlying cosmic acceleration~\cite{Gurrola:2026fqx,Gurrola:2026nks}.

Unfortunately, whereas $w$ is constrained at the percent level by current cosmological probes~\cite{Planck2018,DESI:2025zgx}, $c_s^2$ remains only weakly bounded. Existing probes primarily rely on integrated cosmological effects such as the late-time Integrated Sachs–Wolfe effect and correlations with large-scale structure~\cite{csc1Hannestad:2005ak,csc2de2010measuring,csc3ballesteros2010dark,csc4linton2018variable,csc5bean2004probing,csc6eisenstein2005dark,csc7sergijenko2015sound}, which leave substantial degeneracies among different dark energy models. Therefore, developing new and independent probes of dark energy perturbations is an important goal for both cosmology and fundamental physics.

To this end, high-energy particle colliders provide a complementary window into the fundamental structure of the dark sector. The LHC has an extensive experimental program searching for physics beyond the Standard Model (BSM)~\cite{Shen:2026qxg,Shen:2025nkr,Qureshi:2024cmg,Qureshi:2024naw,Barbosa:2022mmw,Gurrola:2022ssc,Dutta:2022bfq,Cardona:2021ebw,Florez:2021zoo,Florez:2019tqr,Florez:2018ojp,Leonardi:2018jzn,Avila:2018sja,Florez:2017xhf,Florez:2016uob,Florez:2016lwi,Dutta:2015hra,Dutta:2014jda,Dutta:2013gga,Dutta:2012xe,Dutta:2008ge,Arnowitt:2008bz,Delannoy:2013ata}, including extended Higgs sectors~\cite{Branco:2011iw,Craig:2013hca,Gunion:2002zf}, dark matter mediators~\cite{Abercrombie:2015wmb,Abdallah:2015ter,Boveia:2018yeb}, and other new particles that could reveal the microscopic structure of hidden sectors. 
In particular, recent analyses by the CMS and ATLAS collaborations have reported statistically significant excesses in the di-top invariant mass spectrum near the di-top mass threshold~\cite{ATLAS:2026dbe,CMS:2025kzt}. 
These features have been discussed in the literature as possibly arising from QCD effects such as toponium formation or from a new pseudoscalar resonance associated with BSM physics.

A widely used theoretical framework in BSM extensions is the Two-Higgs-Doublet Model supplemented by a pseudoscalar mediator (2HDM+$a$), which can be applied to interpret such excesses ~\cite{Bauer:2017ota,Bauer:2017nlg,No:2015xqa,Arina:2016cqj,arcadi2hdmapheno,Argyropoulos:2024yxo}. This model extends the scalar sector in a renormalizable manner, provides a natural portal between the SM and dark matter (DM), and predicts characteristic collider signatures including $t\bar t$ resonances and missing-energy final states~\cite{Abercrombie:2015wmb,Boveia:2018yeb}. 
However, in essentially all existing collider studies of the 2HDM+$a$ framework, mediator widths and branching ratios are computed purely as functions of particle-physics parameters such as Higgs mixing angles, Yukawa couplings, and dark-matter interactions~\cite{Bauer:2017ota,Bauer:2017nlg}. 
Cosmology enters only indirectly through relic-density considerations~\cite{Arcadi:2017kky}.

Conversely, most dark energy models are built using the language of gravitational or cosmological EFT, without embedding the dark energy sector into a concrete particle-physics framework relevant for collider experiments. This separation between collider physics and dark energy phenomenology leaves largely unexplored the possibility that the physics responsible for cosmic acceleration might also leave detectable signatures in laboratory experiments. Complementary non-gravitational probes of dark energy have begun to be explored, both in terrestrial experiments~\cite{OShea:2024jjw,Yuan:2025twx} and through the scattering of dark energy with visible matter in cosmological settings~\cite{Vagnozzi:2019kvw,Ferlito:2022mok}, while collider searches of the kind we pursue here remain largely unexplored.

The observational case for revisiting the nature of dark energy has become particularly compelling with the recent release of the second DESI data set. DESI DR2 provides precise baryon acoustic oscillation measurements over a broad redshift range and, when combined with CMB and Type-Ia supernova observations, exhibits a preference for a time-evolving dark-energy equation of state over $\Lambda$CDM at a level of approximately $2.8$--$4\sigma$, depending on the external supernova sample and data combination~\cite{DESI:2025zgx,Scherer:2025esj,Silva:2025twg}. Although the statistical significance and robustness of this preference remain under active investigation, the results provide strong motivation to consider dynamical dark energy as a viable possibility rather than merely a theoretical alternative to a cosmological constant. Importantly, and as already noted, measurements of the background equation of state $w(z)$ do not uniquely determine the underlying dark-energy theory. Distinct scalar-field and effective-field-theory realizations can produce similar expansion histories while exhibiting substantially different dynamics for their perturbations. In particular, the dark energy sound speed $c_s^2$ controls the propagation and clustering of perturbations and can differ by orders of magnitude among models with comparable values of $w$. Determining such microphysical properties therefore requires observables beyond the background expansion history. These observation motivates the central question explored in this work: \textit{Can collider measurements of BSM mediator resonances provide a probe of dark energy microphysics?} 
In this paper we develop a unified EFT framework in which the 2HDM+$a$ model is extended by a dynamical dark energy scalar sector with derivative couplings to the pseudoscalar mediator. 
These couplings are not introduced ad hoc, but arise naturally from an approximate shift-symmetry $\phi\to\phi+\text{const}$ that protects the lightness and slow evolution of the dark energy field \cite{Vasilev:2024deshiftsymm, Deffayet:2010dekgb}. Within this framework, interactions with the mediator emerge as a consequence of the underlying structure and can manifest through energy-dependent processes at collider scales. 
We show that these interactions can open a new invisible decay channel $a\to\phi\phi$ and can renormalize the mediator kinetic term in the presence of a dark energy kinetic background. 
Both effects modify the pseudoscalar decay widths, branching ratios, and resonance lineshape relevant for collider searches. 
As a consequence, collider measurements or constraints of BSM mediator properties can become sensitive to parameters that are usually associated with dark energy EFT, including the sound speed of dark energy, defined as pressure perturbations over energy density perturbations of the dark energy field. In this way, symmetry-motivated derivative interactions provide a natural bridge between dark energy microphysics and collider observables, enabling the possibility of probing the physics of cosmic acceleration at colliders.

The remainder of this paper is organized as follows. We begin by constructing the extended EFT Lagrangian and motivating each operator from cosmological EFT considerations. We then derive the kinetic normalization of the pseudoscalar mediator and the resulting rescalings of physical masses and couplings. Next, we compute the partial decay widths, first within the context of canonical quintessence and subsequently extending the calculation to the k-essence scenario, thereby introducing explicit dependence on the sound speed of dark energy. We then analyze distinct scenarios corresponding to different assumptions about the local dark energy kinetic background, followed by a discussion of the collider implications for widths, branching ratios, and kinematic observables. Building on this, we connect the same operators to cosmological perturbation theory and examine the associated theoretical consistency conditions. Finally, we conclude with a proposed program for joint collider--cosmology inference aimed at probing the microphysics of dark energy.

\section{Theoretical Framework}
\label{sec:model}

In order to explore how collider observables may become sensitive to the microphysics of dark energy, we construct a unified effective field theory in which a collider-motivated pseudoscalar mediator sector interacts with a dynamical dark energy scalar field. Our theoretical basis lies in the 2HDM+$a$ model, a framework widely used to interpret heavy-flavor and missing-energy signatures at the LHC~\cite{Bauer:2017ota,Bauer:2017nlg,Abercrombie:2015wmb,Boveia:2018yeb}. In this model, the SM Higgs sector is extended by two Higgs doublets~\cite{Branco:2011iw}, together with an additional pseudoscalar mediator $a$ that mixes with the CP-odd Higgs state after electroweak symmetry breaking~\cite{Bauer:2017ota}. 
This mixing generates a physical pseudoscalar eigenstate that can couple both to SM fermions and to dark-sector particles~\cite{Bauer:2017nlg,No:2015xqa}.

For the collider calculations relevant to this work, we start with couplings of the physical pseudoscalar eigenstate $a$ to top quarks and dark matter. We write its quadratic Lagrangian as 
\begin{equation}
\Lag_{a,0} = \frac12(\partial a)^2 - \frac12 m_a^2 a^2 ,
\label{eq:La0}
\end{equation}
where $m_a$ is a Lagrangian mass parameter for the mediator in the absence of interactions with the dark energy sector. 
The dominant pseudoscalar couplings relevant for collider phenomenology arise from interactions with top quarks and a fermionic dark matter candidate $\chi$, which we write as
\begin{equation}
\Lag_{\rm int}^{(t,\chi)} \supset i g_t\,a\,\bar t\gamma^5 t 
\;+\; 
i g_\chi\,a\,\bar\chi\gamma^5\chi .
\label{eq:Lint_tchi}
\end{equation}
This choice is partially motivated by recent ATLAS and CMS searches for resonances in the $t\bar t$ invariant mass spectrum, which have reported statistically significant excesses near the $2m_t$ threshold~\cite{ATLAS:2026dbe,CMS:2025kzt}. These interactions govern the primary decay channels relevant for di-top and missing-energy searches at the LHC. 
We emphasize that this setup should be regarded as a benchmark scenario chosen for concreteness and experimental relevance. However, the central results of this work are not sensitive to this restriction and can be straightforwardly generalized to cases in which the pseudoscalar mediator couples more broadly to other Standard Model fermions or bosons. In a UV-complete 2HDM+$a$ realization the coupling $g_t$ arises from the Yukawa structure of the two-Higgs-doublet sector and the mixing between pseudoscalar states. For the purposes of the present analysis, however, we treat $g_t$ and $g_\chi$ as effective couplings, to be mapped onto a specific 2HDM+$a$ benchmark at a later stage. The above sector provides a well-defined description of pseudoscalar mediator phenomenology at colliders~\cite{Arina:2016cqj}. 
Our goal is to extend this framework by introducing a dark energy scalar field whose dynamics will influence the mediator properties and therefore modify collider observables.

We introduce a scalar field $\phi$ that plays the role of dark energy and describe it using the general effective-field-theory Lagrangian
\begin{equation}
\Lag_\phi = P(\phi,X),
\qquad
X \equiv -\frac12\,\partial_\mu\phi\,\partial^\mu\phi .
\label{eq:Lag_phi}
\end{equation}
This formulation encompasses a broad class of scalar-field dark energy models~\cite{ArmendarizPicon:1999rj,Garriga:1999vw,Tsujikawa:2013fta}. Canonical quintessence corresponds to the choice $P=X-V(\phi)$~\cite{Ratra:1987rm,Caldwell:1997ii}, while more general k-essence theories allow nonlinear functions of $X$ that modify the kinetic structure of the field~\cite{ArmendarizPicon:2000dh}. This EFT description should be viewed as the lower-energy limit of a more fundamental theory in which the scalar field is responsible for cosmic acceleration at late times.

Within this framework the energy density and pressure associated with the scalar field, identified as $\rho_\phi = T_{00}$ and $p_\phi = \tfrac{1}{3}(T^\mu_{\;\mu} + \rho_\phi)$, are given by
\begin{equation}
p_\phi = P(\phi,X),
\qquad
\rho_\phi = 2X P_X - P ,
\label{eq:rhop}
\end{equation}
where $P_X \equiv \partial P/\partial X \equiv \partial_X P$. The dynamics of perturbations around the cosmological background are governed by the sound speed
\begin{equation}
c_s^2 =
\frac{P_X}{P_X + 2X P_{XX}},
\label{eq:cs2}
\end{equation}
where $P_{XX}= \partial^{2} P/\partial X^{2} = \partial^2_X P$~\cite{ArmendarizPicon:1999rj,Garriga:1999vw}.  

Note that this construction admits $c_s^2$ to deviate significantly from unity ~\cite{Garriga:1999vw,DeDeo:2003te}. This is in contrast with canonical scalar-field models, wherein one typically finds $c_s^2=1$, implying that dark energy remains nearly homogeneous on sub-horizon scales~\cite{Hu1998,Caldwell:1997ii}. 
When $c_s^2\ll1$, dark energy can cluster gravitationally and participate in the growth of structure~\cite{Erickson:2001bq,csc5bean2004probing}, while other scenarios may produce superluminal propagation speeds in certain regimes~\cite{Babichev:2007dw}.
Determining or constraining this quantity therefore provides a direct probe of the microphysical structure of the dark energy sector~\cite{Gurrola:2026fqx,Gurrola:2026nks}.

To connect the dark energy sector with collider phenomenology we consider derivative interactions between the dark energy scalar $\phi$ and the pseudoscalar mediator $a$. 
Derivative couplings are well motivated in many scalar-field theories because approximate shift symmetries suppress direct potential interactions, resulting in the leading interaction involving derivatives of the field. An exact shift symmetry ($\phi \to \phi + \rm constant$) forbids direct potential terms such as mass terms or polynomial potentials, since the Lagrangian cannot depend explicitly on $\phi$ itself \cite{Vasilev:2024deshiftsymm, Deffayet:2010dekgb}. 

In this context, direct potential interactions would generically introduce large mass scales or rapid evolution for the scalar field. 
Such interactions are typically disfavored in dark energy model building, since they can spoil the slow-roll or slowly varying behavior required for cosmic acceleration, and can lead to strong constraints from laboratory and astrophysical tests \cite{Berbig:2025DESIquint, Chiang:supergravquint, Finelli:2018shiftsymmcosm}. By contrast, derivative interactions such as $a\,(\partial\phi)^2$
are compatible with approximate shift symmetry and therefore arise naturally in effective field theories of light scalar degrees of freedom. 
This structure is familiar from Goldstone bosons and other modes associated with spontaneously broken symmetries, where the leading interactions are derivative-suppressed \cite{Berbig:2025DESIquint, Kaneta:2024pngbinflation}. 
In the context of dark energy, similar symmetry arguments are often invoked to ensure radiative stability of the scalar sector and to protect the small mass scale associated with cosmic acceleration.

In the present work, dark energy derivative couplings occur with the physical pseudoscalar $a$ and can thus leave imprints on collider observables depending on the energy scale. The leading operator that directly connects the mediator to the dark energy sector is a dimension-five derivative interaction,
\begin{equation}
\Lag_{a\phi\phi} =
\frac{C_{5}}{\Lambda_5}\,a\,\partial_\mu\phi\,\partial^\mu\phi ,
\label{eq:dim5}
\end{equation}
where $\Lambda_5$ represents the characteristic mass scale of the heavy degrees of freedom that generate this operator upon being integrated out, and the coefficient $C_5$ is a dimensionless Wilson coefficient that encodes the strength of the underlying UV interactions. 
The ratio $C_{5}/\Lambda_{5}$, with units $\text{GeV}^{-1}$, therefore plays the role of the effective coupling controlling the interaction at low energies. This term represents a trilinear interaction in which the mediator $a$ can emit or absorb excitations of the scalar field $\phi$. 
In practical terms this operator opens a new invisible decay channel, 
$a \rightarrow \phi\phi$, 
provided that excitations of $\phi$ are stable on collider timescales. 
We note that the dimension-five interaction in Eq.~\ref{eq:dim5} is CP violating under the standard CP assignments of the 2HDM+$a$ framework. In particular, $a$ is CP odd, while $\phi$ is taken to be a scalar and $(\partial\phi)^2$ is CP even, such that $a(\partial\phi)^2$ is CP odd. Accordingly, $C_5/\Lambda_5$ should be interpreted as a CP-violating portal between the pseudoscalar mediator and the dark energy sector. Such an operator may arise after integrating out heavy states whose interactions or mixing violate CP, including scenarios with explicit or spontaneous CP violation in the ultraviolet theory. We remain agnostic about this UV completion and treat $C_5$ as a phenomenological Wilson coefficient. The presence of this CP-violating interaction does not, by itself, imply a direct tree-level contribution to conventional low-energy CP observables, since Eq.~\ref{eq:dim5} contains no Standard Model fields. CP violation may nevertheless be communicated to the visible sector through the mediator couplings, generating electric-dipole-moment or related observables at loop level. The resulting constraints depend on the mediator couplings, the heavy states responsible for generating Eq.~(6), and the detailed UV completion. A dedicated analysis of such constraints is therefore beyond the scope of the present collider EFT study, but would be required in a fully specified ultraviolet realization.

Physically, the dimension-five interaction allows energy stored in the mediator field to be converted into propagating fluctuations of the dark energy sector. The interpretation of this interaction can be understood by analogy with other systems in physics. 
For example, in electromagnetism a scalar field coupled through an operator such as $\phi F_{\mu\nu}F^{\mu\nu}$ can source or absorb electromagnetic radiation~\cite{Sikivie:1983ip,Kim:1986ax,Marsh:2015xka}. 
Similarly, in condensed-matter systems, lattice deformations can couple to phonons, allowing energy to be transferred between background strains and propagating excitations~\cite{Ziman:1960,Kittel:2004}. 
In the present case, the pseudoscalar mediator interacts with background kinetic excitations of the dark energy field in an analogous manner.

A second interaction arises at dimension eight,
\begin{equation}
\Lag_{\rm kinmix} =
\frac{C_8}{\Lambda_8^4}
(\partial\phi)^2(\partial a)^2 ,
\label{eq:dim8}
\end{equation}
where $\Lambda_8$,  $C_8$, and the ratio $C_8/\Lambda_8^4$ play the roles of $\Lambda_5$, $C_5$ and $C_{5}/\Lambda_{5}$ in the dimension-five case (here, the ratio has units $\text{GeV}^{-4}$). Unlike the previous operator, this term does not correspond directly to a decay vertex. 
Instead, it modifies the kinetic structure of the mediator field when the dark energy field possesses a nonzero background gradient.

In a cosmological setting the dark energy field generically evolves in time, implying that the background value of $X$ is nonzero. 
In such a background the operator in Eq.~(\ref{eq:dim8}) induces an effective rescaling of the mediator kinetic term,
$(\partial a)^2 \rightarrow \frac{(\partial a)^2}{Z_a}$ , 
where the parameter $Z_a$ depends on the background value of $(\partial\phi)^2$. 
This effect alters the normalization, propagation, and dispersion properties of the pseudoscalar field. Conceptually, this operator can be interpreted in a similar vein to the propagation of particles in a medium. 
For example, photons traveling through an electron plasma acquire an effective dispersion relation because the background plasma modifies their kinetic response~\cite{Braaten:1990it,Kapusta:2006pm}. Similarly, phonons in condensed-matter systems propagate differently depending on the background properties of the lattice~\cite{Ziman:1960,Kittel:2004}. 
In the present case, the cosmological dark energy background acts as a dynamical medium that modifies the effective kinetic properties of the pseudoscalar mediator.

From a collider perspective, this interaction implies that the wavefunction normalization of the mediator can depend on the dark energy background. 
Consequently, production rates, decay widths, and kinematic distributions of the mediator may become sensitive to parameters associated with the dark energy sector.

The derivative structure of Eqs.~\ref{eq:dim5} and~\ref{eq:dim8} is also important for the radiative stability of the dark-energy sector. A generic quintessence scalar is subject to the well-known naturalness problem that its mass must remain of order the present Hubble scale, $m_\phi\sim H_0$, despite the presence of much heavier degrees of freedom. In the approximate shift-symmetric setup adopted here, $\phi$ enters the mediator interactions only through derivatives. In the exact symmetry limit, these interactions therefore do not generate an arbitrary potential $V(\phi)$ or an unsuppressed mass term for $\phi$. The scales $\Lambda_5$ and $\Lambda_8$ should accordingly be interpreted as matching scales associated with the heavy physics generating the derivative portal, rather than as the scale responsible for the quintessence potential itself. This symmetry structure ameliorates the radiative sensitivity of the ultralight scalar to the collider sector, but does not by itself solve the full UV naturalness problem of quintessence.

The approximate nature of the shift symmetry nevertheless remains an essential model-building assumption. Explicit symmetry breaking is required to generate the small potential responsible for the cosmological evolution of $\phi$, and a complete ultraviolet realization must ensure that radiative and gravitational corrections associated with this breaking do not destabilize the required hierarchy. Addressing this longstanding quintessence naturalness problem lies beyond the scope of the collider EFT considered here. Our assumption is therefore that the derivative portal is embedded in a technically controlled dark energy sector in which the required small symmetry-breaking parameters are radiatively stable.

To ensure stability of dark energy scalar excitations, we impose a discrete $\mathbb{Z}_2$ symmetry under which $\phi \;\to\; -\phi$. Under this symmetry, all operators involving an odd power of $\phi$ fields are forbidden. 
In particular, linear couplings of $\phi$ to SM fields or to other sectors are absent, preventing the decay of a single $\phi$ quantum into lighter states. 
As a result, $\phi$ excitations produced in processes such as $a\to\phi\phi$ are stable on detector timescales and manifest as missing momentum at colliders. Note that this $\mathbb{Z}_2$ symmetry does not prevent a time-dependent background configuration $\bar\phi(t)$, which continues to evolve according to $P(\phi,X)$ and can drive cosmic acceleration. 

The stability of the dark matter candidate $\chi$ is, in general, an independent assumption of the 2HDM+$a$ framework and may arise from a separate symmetry, often another $\mathbb{Z}_2$ under which $\chi$ is odd. In the setup considered here, $\chi$ is taken to be stable and thus it contributes to missing-energy signatures, but its stabilization need not be tied to the $\phi$ sector. Nevertheless, if desired, one could embed both $\phi$ and $\chi$ within a common discrete symmetry structure, leading to a unified origin of stability for invisible final states.

In any case, collecting the relevant terms, the effective Lagrangian governing the interactions of the mediator and the dark energy sector is
\begin{multline}
\Lag =
\frac12(\partial a)^2 - \frac12 m_a^2 a^2
+ i g_t\,a\,\bar t\gamma^5 t
+ i g_\chi\,a\,\bar\chi\gamma^5\chi \\
+ P(\phi,X)
+ \frac{C_5}{\Lambda_5}a(\partial\phi)^2
+ \frac{C_8}{\Lambda_8^4}(\partial\phi)^2(\partial a)^2,
\label{eq:fullLag}
\end{multline}
providing a minimal framework in which collider observables associated with the pseudoscalar mediator can become sensitive to the kinetic structure of the dark energy sector. 

\section{Mediator Kinetic Normalization and Physical Parameters}
\label{sec:kinmix}

Unlike the dimension--5 interaction~\eqref{eq:dim5}, that opens a new decay channel, the dimension--8 operator~\eqref{eq:dim8} modifies the kinetic structure of the mediator in the presence of a nontrivial dark energy kinetic background.

It is convenient to express the interaction in terms of the scalar quantity $X \equiv -\frac12 \partial_\mu\phi \partial^\mu\phi$. 
Using the identity $(\partial\phi)^2 \equiv \partial_\mu\phi\,\partial^\mu\phi = -2X$, the kinetic mixing operator becomes
\begin{align}
\Lag_{\rm kinmix}
= \frac{C_8}{\Lambda_8^4}(\partial\phi)^2(\partial a)^2
= \frac{-2XC_8}{\Lambda_8^4}(\partial a)^2 .
\label{eq:kinmixX}
\end{align}
We now combine this contribution with the canonical kinetic term of the mediator to obtain
\begin{equation}
\Lag_{a,\rm kin} =
\frac12(\partial a)^2
-
\frac{2XC_8}{\Lambda_8^4}(\partial a)^2=\frac12
\left(
1
-
\frac{4XC_8}{\Lambda_8^4}
\right)
(\partial a)^2 .
\end{equation}
This expression motivates the definition of a dimensionless parameter
\begin{equation}
Z_a \equiv
1 - 4X\frac{C_8}{\Lambda_8^4}.
\label{eq:Za_def}
\end{equation}
The quadratic Lagrangian governing the mediator field therefore becomes
\begin{equation}
\Lag_{a,\rm quad}
=
\frac12 Z_a (\partial a)^2
-
\frac12 m_a^2 a^2 .
\label{eq:Laquad}
\end{equation}

The parameter $Z_a$ measures how the kinetic normalization of the mediator $a$ is modified by the background kinetic energy of the dark energy field. 
When $X=0$ the standard vacuum normalization is recovered and $Z_a=1$. 
However, in a scenario where the dark energy field evolves in time, $X$ acquires a nonzero expectation value and the effective normalization of the mediator field is shifted ($Z_a\neq 1$). This in turn alters effective propagation properties of $a$. 
We note that appreciable deviations $Z_a\neq 1$ should be interpreted phenomenologically as allowing for a local value of $X$ in the terrestrial environment that differs from its homogeneous cosmological value, as may occur in nonlinear scalar EFTs whose solutions are sensitive to the surrounding matter distribution or admit spatially varying field profiles. The present analysis does not specify a mechanism that generates such a configuration on Earth. Determining the local value of $X$ in a particular model requires solving the corresponding nonlinear scalar-field equations in the presence of the relevant matter distribution and then checking collider/laboratory and Solar-System constraints, which lies beyond the scope of the collider EFT considered here. Accordingly, $Z_a$ is treated as an effective parameter encoding the collider response to a possible local kinetic background, rather than as a prediction of the $Z_a\neq1$ value produced by a given terrestrial environment. We regard $Z_a$ as a useful phenomenological parameter because it allows the collider analysis to remain agnostic about the microscopic mechanism that establishes the local scalar background.

It is also important to note that a nonzero background value of $X$ does not, by itself, imply the existence of a fifth force or an equivalence-principle violation. In the EFT considered here, the dark-energy field $\phi$ has no direct coupling to Standard Model matter. The $\mathbb{Z}_2$ symmetry forbids direct linear couplings of $\phi$ to Standard Model matter, and the visible sector communicates with $\phi$ only indirectly through the pseudoscalar mediator and the derivative operators. The leading interactions with the visible sector arise indirectly through the pseudoscalar mediator $a$, through the operators $\frac{c}{\Lambda}a(\partial\phi)^2$ and $\frac{1}{\Lambda_8^4}(\partial\phi)^2(\partial a)^2$. Consequently, the existence of a background kinetic energy $X$ does not automatically generate a macroscopic force between ordinary matter sources. Conventional fifth-force and equivalence-principle constraints become directly applicable only after specifying an additional coupling of $\phi$ to matter, or a nonminimal coupling to gravity, neither of which is assumed in the collider EFT studied here. This distinction is important because fifth-force constraints on scalar-tensor and kinetically screened theories generally constrain the scalar's coupling to matter in addition to its nonlinear kinetic structure.

For collider processes the relevant spacetime region is microscopic compared to cosmological length and time scales. 
We therefore adopt the standard EFT assumption that the background value of $X$ is approximately constant over the spacetime region probed by a collider event. 
Under this separation-of-scales assumption, $Z_a$ can be treated as a constant parameter in the calculation of mediator production and decay.

The kinetic term in Eq.~(\ref{eq:Laquad}) is not canonically normalized when $Z_a \neq 1$. 
To obtain the physical propagating field we introduce a canonically normalized field $a_c$ defined by $a_c \equiv \sqrt{Z_a}\,a$, or 
equivalently, $a = a_{c}/\sqrt{Z_a}$.
Since $Z_a$ is constant, derivatives act only on the field itself:
\begin{align}
\partial_\mu a
=
\partial_\mu
\left(
\frac{a_c}{\sqrt{Z_a}}
\right)
=
\frac{1}{\sqrt{Z_a}}\,\partial_\mu a_c .
\end{align}
Substituting this relation into Eq.~(\ref{eq:Laquad}) yields
\begin{align}
\Lag_{a,\rm quad}
&=
\frac12 Z_a
\left(
\frac{1}{Z_a}
(\partial a_c)^2
\right)
-
\frac12 m_a^2
\left(
\frac{a_c}{\sqrt{Z_a}}
\right)^2
\nonumber\\
&=
\frac12(\partial a_c)^2
-
\frac12
\frac{m_a^2}{Z_a}
a_c^2 .
\end{align}
We therefore identify the physical mediator mass as 
\begin{equation}
m_{a,\rm phys} =
\frac{m_a}{\sqrt{Z_a}},
\label{eq:ma_phys}
\end{equation}
which demonstrates that the dark energy kinetic background induces a shift in the physical mass of the mediator. From a collider perspective this effect corresponds to a background-dependent normalization of the mediator mass.

We note that the kinetic term of a healthy propagating scalar field must have the correct sign in order to avoid ghost instabilities. 
This requirement implies $Z_a > 0$, which in turn requires that $X < \Lambda_8^4/4C_8$ via Eq.~(\ref{eq:Za_def}). This inequality represents a consistency condition on the allowed range of the dark energy kinetic background relative to the cutoff scale of the effective theory.

An important consequence of canonical normalization is that all interactions linear in the mediator field acquire a rescaling factor. Consider for example the coupling to top quarks,
$i g_t a \bar t \gamma^5 t$. Expressing the interaction in terms of the canonical field $a_c$ gives
\begin{align}
i g_t a \bar t \gamma^5 t
&=
i g_t
\left(
\frac{a_c}{\sqrt{Z_a}}
\right)
\bar t \gamma^5 t .
\end{align}
Therefore, we define the physical coupling
\begin{equation}
g_{t,\rm phys} =
\frac{g_t}{\sqrt{Z_a}} .
\label{eq:gt_phys}
\end{equation}
The same rescaling applies to the dark-matter coupling,
\begin{equation}
g_{\chi,\rm phys} =
\frac{g_\chi}{\sqrt{Z_a}}.
\label{eq:gx_phys}
\end{equation}
The derivative interaction with the dark energy field transforms in an analogous way:
\begin{align}
\frac{C_5}{\Lambda_5}a(\partial\phi)^2
&=
\frac{C_5}{\Lambda_5}
\left(
\frac{a_c}{\sqrt{Z_a}}
\right)
(\partial\phi)^2 .
\end{align}
Defining the physical coefficient
\begin{equation}
C_{5,\rm phys} =
\frac{C_5}{\sqrt{Z_a}},
\end{equation}
the interaction becomes
\begin{equation}
\Lag_{a\phi\phi}
=
\frac{C_{5,\rm phys}}{\Lambda_5}
a_c(\partial\phi)^2 .
\label{eq:aphiphiLangrangian}
\end{equation}
These results demonstrate that the kinetic mixing operator produces correlated shifts in the BSM mediator mass and its interaction strengths. 
Consequently, collider observables such as production cross sections, decay widths, and resonance lineshapes can become sensitive to the dark energy kinetic background encoded in the parameter $Z_a$. 

\section{Mediator Partial Widths with Derivative Dark Energy Couplings}
\label{sec:widths}

In this section we derive the mediator partial widths under a set of simplifying but well-defined assumptions that closely parallel standard calculations in particle physics. We take the dark energy sector to be in the \emph{canonical quintessence} limit, $P(\phi,X)=X - V(\phi)$, for which the field $\phi$ is canonically normalized and Lorentz invariance remains manifest at the level of fluctuations. 
Expanding around a background configuration and keeping the quadratic terms for excitations, one obtains the standard relativistic scalar form
\begin{equation}
\Lag_{\phi,0}\supset \frac12(\partial\phi)^2-\frac12 m_\phi^2\phi^2,
\label{eq:Lphi0_canonical}
\end{equation}
where $m_\phi^2 \equiv V_{\phi\phi}$ evaluated on the background.
Under these assumptions, excitations of $\phi$ propagate with the Lorentz-invariant dispersion relation $\omega^2=\bm{k}^2+m_\phi^2$, and the field $\phi$ creates properly normalized asymptotic one-particle states. 
As a result, decay widths can be computed using the familiar two-body phase-space measure and traditional S-matrix expressions, exactly as one would for decays into ordinary scalar particles.

Conceptually, this setup corresponds to the intuitive picture in which the mediator decays into two freely propagating particles whose energy, momentum, and normalization follow the same rules as those of standard relativistic quanta. 
The calculation performed here therefore serves as a transparent baseline: it isolates the role of the derivative cubic interaction and allows one to track explicitly how the mediator mass, the coupling scale $\Lambda_5$, and the kinetic normalization parameter $Z_a$ enter the decay rate. 
These assumptions do not capture the most general behavior expected of a dynamical dark energy sector. 
In particular, for a generic $P(\phi,X)$ theory the fluctuations of $\phi$ need not be canonically normalized, nor do they necessarily propagate at the speed of light. 
In later sections we relax these assumptions and perform a more general calculation in which the physical decay width is expressed in terms of canonically normalized perturbations and a nontrivial dispersion relation characterized by the dark energy sound speed $c_s^2$. 

We compute the partial widths relevant for collider di-top and invisible searches, 
$a \to t\bar t$, 
$a \to \chi\bar\chi$, 
$a \to \phi\phi$, 
working throughout with the canonically normalized mediator field $a_c$ and expressing results in terms of the original set of parameters
$(m_a,g_t,g_\chi,C_5/\Lambda_5,C_8/\Lambda_8,X)$ via the kinetic normalization $Z_a$ introduced in the previous section. 
The physical mediator mass will be simply denoted by $m$, such that $m \equiv m_{a,\rm phys}=\frac{m_a}{\sqrt{Z_a}}$. 

When $Z_a\simeq 1$, the mediator is essentially canonically normalized at collider scales and the dominant effect of the dark energy sector is the opening of the new invisible channel through the dimension--5 operator. This would be the case in the cosmological-background-only expectation, since $X$ (set by the homogeneous dark energy kinetic density) would be extraordinarily small in collider units and thus $4XC_8/\Lambda_8^4\ll1$ for any $\Lambda_8$ near or above the electroweak scale.
Meanwhile, when $Z_a$ differs appreciably from unity, the mediator wavefunction normalization, physical mass, and all linear couplings are simultaneously rescaled, leading to correlated shifts across visible and invisible channels. This scenario can be interpreted as a parameterization of nontrivial local kinetic backgrounds, for example in environments where gradients are enhanced or where the EFT admits localized configurations. In such cases, the term $4XC_8/\Lambda_8^4$ need not be much less than $1$, and thus we may have $0<Z_a<1$. Physically, the dimension--8 operator makes the mediator propagate in an effective medium set by the local value of $(\partial\phi)^2$. 

\subsection{Decay $a\to\phi\phi$ from the derivative cubic}
\label{subsec:width_phi_nosubsec}

We begin by computing the partial decay width of the pseudoscalar mediator into two dark energy quanta,
$a_c \to \phi\phi,$ arising from the derivative interaction introduced in the previous section. 
In the canonical quintessence limit, the field $\phi$ behaves as a standard relativistic scalar, and the calculation can be performed using familiar S-matrix techniques. From Eq.~\eqref{eq:aphiphiLangrangian}, the interaction relevant for this decay is
\begin{equation}
\Lag \supset \frac{C_5}{\Lambda_5\sqrt{Z_a}}\,a_c\,\partial_\mu\phi\,\partial^\mu\phi .
\label{eq:Lint_a_phiphi}
\end{equation}
This operator describes a trilinear interaction in which the mediator couples to the local kinetic energy density of the $\phi$ field.
As a result, the decay amplitude is proportional to the momenta carried by the outgoing particles.

\ifcalcsinbody
To calculate the decay amplitude, we consider the decay process
\begin{equation}
a_c(p)\to \phi(p_1)\,\phi(p_2),
\qquad p=p_1+p_2 ,
\label{eq:decay_def_phiphi}
\end{equation}
where $p$ is the four-momentum of the mediator and $p_1$, $p_2$ are the momenta of the final-state particles. 
In momentum space, each derivative acting on an external $\phi$ field contributes a factor of $i p_\mu$. 
Applying this rule to Eq.~\eqref{eq:Lint_a_phiphi}, the tree-level squared matrix element becomes

\begin{equation}
|\mathcal{M}|^2
=
\left(\frac{C_5}{\Lambda_5\sqrt{Z_a}}\right)^2 (p_1\cdot p_2)^2,
\label{eq:M2_phiphi_1}
\end{equation}
from which we can see that the interaction is momentum-suppressed. To proceed with the evaluation of kinematic invariants, we express the Lorentz-invariant product $p_1\cdot p_2$ in terms of physical masses. 
Using the identity
\begin{equation}
p^2=(p_1+p_2)^2=p_1^2+p_2^2+2p_1\cdot p_2,
\label{eq:p2_identity}
\end{equation}
and imposing the on-shell conditions $p^2=m^2$ and 
$p_1^2=p_2^2=m_\phi^2$, we obtain
\begin{equation}
m^2 = 2m_\phi^2 + 2p_1\cdot p_2,
\end{equation}
which can be rearranged to give
\begin{equation}
p_1\cdot p_2=\frac{m^2-2m_\phi^2}{2}.
\label{eq:p1p2_eval}
\end{equation}
Substituting this expression into Eq.~\eqref{eq:M2_phiphi_1}, the squared amplitude becomes
\begin{equation}
|\mathcal{M}|^2
=
\frac{C_5^2}{\Lambda_5^2 Z_a}\,
\frac{(m^2-2m_\phi^2)^2}{4}.
\label{eq:M2_phiphi}
\end{equation}
This form makes the dependence on both the mediator mass and the dark energy scalar mass explicit.

The partial decay width is then computed using the standard two-body phase-space formula,
\begin{equation}
\Gamma=\frac{1}{2m}\,\frac{1}{2!}\int \dd\Phi_2\,|\mathcal{M}|^2,
\label{eq:Gamma_def}
\end{equation}
where the factor of $1/2!$ accounts for the identical $\phi$ particles in the final state. 
The integrated two-body phase space is given by
\begin{equation}
\int\dd\Phi_2 = \frac{1}{8\pi}\beta_\phi,
\qquad
\beta_\phi \equiv \sqrt{1-\frac{4m_\phi^2}{m^2}} ,
\label{eq:Phi2}
\end{equation}
where $\beta_\phi$ represents the velocity of the final-state particles in the rest frame of the decaying mediator. Substituting Eq.~\eqref{eq:Phi2} into Eq.~\eqref{eq:Gamma_def}, we obtain
\begin{equation}
\Gamma(a_c\to\phi\phi)
=
\frac{|\mathcal{M}|^2}{32\pi m}\,\beta_\phi,
\end{equation}
and finally, inserting Eq.~\eqref{eq:M2_phiphi} yields
\begin{equation}
\Gamma(a_c\to\phi\phi)
=
\frac{C_5^2}{128\pi}\,
\frac{(m^2-2m_\phi^2)^2}{\Lambda_5^2 Z_a\,m}\,
\sqrt{1-\frac{4m_\phi^2}{m^2}} .
\label{eq:Gamma_phiphi_phys}
\end{equation}

It is useful to express the result in terms of the original Lagrangian mass parameter $m_a$ and the kinetic normalization $Z_a$. 
Using the relation
$m^2=\frac{m_a^2}{Z_a}$, we obtain
\begin{equation}
\begin{split}
\Gamma(a\to\phi\phi)
&=
\frac{C_5^2}{128\pi}\,
\frac{(m_a^2-2Z_a m_\phi^2)^2}{\Lambda_5^2\,m_a}\,\\
&\times Z_a^{-5/2}\,
\sqrt{1-\frac{4Z_a m_\phi^2}{m_a^2}} .
\end{split}
\label{eq:Gamma_phiphi_maZa}
\end{equation}
This expression highlights the nontrivial scaling with $Z_a$, which arises from the combined effects of field normalization, coupling rescaling, and phase space.
\else

\fi

In many phenomenologically relevant scenarios, the dark energy scalar is extremely light compared to collider scales, $m_\phi \ll m_a$~\cite{Ratra:1987rm,Caldwell:1997ii,Tsujikawa:2013fta}. 
In this limit, the phase-space suppression becomes negligible and the expression simplifies considerably:
\begin{equation}
\Gamma(a\to\phi\phi)\;\xrightarrow[m_\phi\to 0]{}\;
\frac{C_5^2}{128\pi}\,\frac{m_a^3}{\Lambda_5^2}\,Z_a^{-5/2}.
\label{eq:Gamma_phiphi_light}
\end{equation}
We see explicitly that the invisible decay width scales as $m_a^3$ and is strongly enhanced for small $Z_a$, reflecting the sensitivity of the decay to the underlying dark energy background through kinetic mixing.

\subsection{Decay $a\to t\bar t$ and $a\to\chi\bar\chi$}
\label{subsec:widths_fermions_nosubsec}

We now summarize the partial decay widths of the pseudoscalar mediator into fermionic final states, focusing on the top-quark channel and the dark matter channel. 
These decay modes are standard in the literature for pseudoscalar mediators and serve as the primary visible and invisible benchmarks against which the new dark energy channel will be compared~\cite{Bauer:2017ota,Bauer:2017nlg,Arina:2016cqj,Abdallah:2015ter}.

In terms of the canonically normalized mediator field $a_c$, the relevant interaction Lagrangian is given by
\begin{equation}
\Lag \supset i g_t^{(\rm phys)} a_c \bar t\gamma^5 t
\;+\;
i g_\chi^{(\rm phys)} a_c \bar\chi\gamma^5 \chi,
\,\,\,\,\,
g_{t,\chi}^{(\rm phys)}=\frac{g_{t,\chi}}{\sqrt{Z_a}} .
\end{equation}
As discussed previously, the factor of $Z_a^{-1/2}$ arises from the canonical normalization of the mediator field, and implies that all fermionic couplings inherit a dependence on the dark energy kinetic background.

The decay width of a pseudoscalar particle into a Dirac fermion pair $f\bar f$ is well known and can be written in the compact form
\begin{equation}
\Gamma(a_c\to f\bar f)
=
N_c\,\frac{\left(g_f^{(\rm phys)}\right)^2\,m}{8\pi}\,
\sqrt{1-\frac{4m_f^2}{m^2}},
\label{eq:Gamma_ff_standard}
\end{equation}
where $N_c$ is the number of colors of the fermion, $m$ is the physical mass of the mediator, and the square-root factor encodes the usual two-body phase-space suppression near threshold.

Applying this general result to the top-quark final state, and recalling that $N_c=3$ for quarks, we obtain
\begin{equation}
\Gamma(a_c\to t\bar t)
=
3\,\frac{\left(g_t^{(\rm phys)}\right)^2\,m}{8\pi}\,
\sqrt{1-\frac{4m_t^2}{m^2}}.
\end{equation}
Substituting the relations $g_t^{(\rm phys)}=g_t/\sqrt{Z_a}$ and $m=m_a/\sqrt{Z_a}$, derived in the previous section, leads to the expression
\begin{equation}
\Gamma(a\to t\bar t)
=
3\,\frac{g_t^2\,m_a}{8\pi}\,
Z_a^{-3/2}\,
\sqrt{1-\frac{4Z_a m_t^2}{m_a^2}} ,
\label{eq:Gamma_tt_maZa}
\end{equation}
which makes explicit how the kinetic normalization parameter $Z_a$ modifies both the overall normalization and the kinematic threshold of the decay.

An entirely analogous calculation applies to the dark matter channel. 
Assuming a Dirac fermion $\chi$ with pseudoscalar coupling $g_\chi$, the decay width is given by
\begin{equation}
\Gamma(a_c\to \chi\bar\chi)
=
\frac{\left(g_\chi^{(\rm phys)}\right)^2\,m}{8\pi}\,
\sqrt{1-\frac{4m_\chi^2}{m^2}}.
\end{equation}
Expressing this in terms of the original Lagrangian parameters yields
\begin{equation}
\Gamma(a\to\chi\bar\chi)
=
\frac{g_\chi^2\,m_a}{8\pi}\,
Z_a^{-3/2}\,
\sqrt{1-\frac{4Z_a m_\chi^2}{m_a^2}} .
\label{eq:Gamma_chichi_maZa}
\end{equation}

For completeness, we note that if the dark matter particle is Majorana rather than Dirac, the decay width in Eq.~\eqref{eq:Gamma_chichi_maZa} is reduced by a factor of $1/2$ due to identical-particle statistics. We emphasize that the presence of the dark energy sector introduces a nontrivial and correlated dependence on $Z_a$.

\subsection{Total width, branching ratios, and the role of $Z_a$}
\label{subsec:total_nosubsec}

Having derived the individual partial widths, we now combine these results to obtain the total decay width and branching fractions of the pseudoscalar mediator. 
These quantities are directly related to collider observables, such as resonance lineshapes, signal rates, and missing-energy signatures, and therefore provide the primary link between the underlying Lagrangian and experimental measurements.

In the dominant-channels approximation emphasized in this work, the total decay width can be written as
\begin{equation}
\Gamma_{\rm tot}
\simeq
\Gamma(a\to t\bar t) + \Gamma(a\to\chi\bar\chi) + \Gamma(a\to\phi\phi).
\label{eq:Gamma_tot}
\end{equation}
This expression captures the leading visible and invisible decay modes relevant for LHC searches in the parameter space of interest. From the total width, the branching fraction into a given final state $i$ is defined in the usual way as
\begin{equation}
{\rm BR}(a\to i)=\frac{\Gamma(a\to i)}{\Gamma_{\rm tot}},
\qquad
i\in\{t\bar t,\chi\bar\chi,\phi\phi\}.
\label{eq:BR_def}
\end{equation}
It is these branching ratios, together with the total width, that determine the relative strength of different collider signatures, including di-top resonances and missing-energy final states.

\subsection{Role of $Z_a$ and physical interpretation}

A central result of this framework is that all partial widths inherit a dependence on the kinetic-normalization parameter $Z_a$, which encodes the effect of the dark energy background on the mediator sector. As discussed previously, $Z_a$ modifies both the physical mass and the effective couplings of the mediator, and therefore enters all decay rates in a correlated manner.

In the regime where $Z_a \simeq 1$, corresponding to the expectation that the LHC probes only the homogeneous cosmological dark energy background, the mediator is effectively canonically normalized.
In this limit, the kinematic thresholds and visible decay rates are close to their standard values, and the primary new effect arises from the additional invisible decay channel $a\to\phi\phi$. 
Even in this conservative scenario, the presence of this channel can significantly alter collider observables by increasing the total width and suppressing the visible branching fractions. 
In particular, a nonzero $\Gamma(a\to\phi\phi)$ leads to a reduction in ${\rm BR}(a\to t\bar t)$ and a corresponding enhancement of missing-energy signatures, providing a direct handle on the derivative coupling.

When $Z_a \neq 1$, the mediator propagates in an effective medium determined by the local kinetic background of the dark energy field.
The physical mass is shifted according to $m=m_a/\sqrt{Z_a}$, which modifies the available phase space for all decay channels. 
At the same time, the effective couplings are rescaled, leading to a nontrivial dependence of the widths on $Z_a$.

Importantly, the different decay channels scale differently with $Z_a$:
\begin{equation}
\Gamma(a\to\phi\phi)\propto Z_a^{-5/2},
\qquad
\Gamma(a\to t\bar t,\chi\bar\chi)\propto Z_a^{-3/2}.
\end{equation}
This difference in scaling implies that even moderate deviations from $Z_a=1$ can induce sizable and correlated changes in the branching ratios. 
In particular, the dark energy channel is more sensitive to the case of $Z_a<1$ than the fermionic channels, making it a powerful probe of dark energy effects.

\subsection{Implications for collider searches}

The phenomenological impact of these effects depends sensitively on the mediator mass. 
In the region $m_a \simeq 350$--$400~{\rm GeV}$, the decay $a\to t\bar t$ lies close to threshold, and the phase-space factor $\sqrt{1-(4Z_a m_t^2 / m_a^2)}$ plays a crucial role. 
In this regime, even small shifts in $Z_a$ can lead to large changes in the visible width, since the decay is highly sensitive to the available phase space. 
As a result, the interplay between the threshold suppression and the additional invisible channel can significantly distort the di-top invariant-mass distribution, providing a particularly sensitive probe of the underlying dynamics.

Away from the threshold region, for mediator masses well above $2m_t$, the phase-space suppression becomes negligible and the fermionic widths scale approximately linearly with $m_a$. 
In this regime, the relative importance of the dark energy channel is controlled primarily by the ratio
\begin{equation}
\frac{\Gamma(a\to\phi\phi)}{\Gamma(a\to t\bar t)}
\;\sim\;
\frac{C_5^2}{\Lambda_5^2}\frac{m_a^2}{g_t^2}\,Z_a^{-1}.
\end{equation}
This scaling shows that at higher masses, the invisible decay mode can become increasingly important, particularly for moderately large values of the derivative coupling $C_5/\Lambda_5$.

As a consequence, even in the absence of a near-threshold enhancement, the dark energy channel can lead to observable deviations in both the total width and the branching ratios. 
These effects would manifest as a broadening of the resonance and a redistribution of signal strength between visible and invisible final states, which could be probed at colliders such as the LHC through a combination of di-top searches and missing-energy analyses. Importantly, these effects are not restricted to a narrow mass window, but persist across a wide range of mediator masses. 
This implies that the framework developed in this work can be tested not only in the context of current di-top excesses, but also in future searches for heavier BSM pseudoscalar resonances.

The results presented here demonstrate that the parameter $Z_a$ provides a direct bridge between dark energy physics and collider observables. In the next section, we move beyond the canonical quintessence limit and consider a more general $P(\phi,X)$ theory. 
This allows us to express the decay width in terms of the sound speed of dark energy $c_s^2$, thereby making explicit the connection between collider measurements and the microphysics of cosmic acceleration. 

\section{Collider Dependence on the Dark Energy Sound Speed}
\label{sec:cs2_colliders}

In the preceding sections, we constructed an EFT extension of the 2HDM+$a$ framework in which a dynamical dark energy scalar $\phi$ couples derivatively to the pseudoscalar mediator $a$. 
Two operators were shown to play distinct and complementary roles: the dimension-5 interaction $a(\partial\phi)^2/\Lambda_5$ opens the invisible decay channel $a\to\phi\phi$, while the dimension-8 kinetic-mixing operator $(\partial\phi)^2(\partial a)^2/\Lambda_8^4$ renormalizes the mediator kinetic term in the presence of a nonzero dark energy kinetic background $X\equiv -\frac12(\partial\phi)^2$. 
Up to this point, the analysis has been performed under the simplifying assumption that the dark energy sector behaves as a canonically normalized relativistic scalar.

We now move beyond this canonical quintessence limit 
and present a central result of this work: collider observables, including the mediator width, branching fractions, and resonance kinematics, can acquire a direct and calculable dependence on the \emph{sound speed of dark energy}, $c_s^2$, in more general settings. 
The origin of this dependence is subtle but physically transparent. 
\emph{While the interaction Lagrangian is written in terms of the field $\phi$, the particles produced at colliders correspond to fluctuations around a cosmological background}, and these fluctuations are governed by a nontrivial quadratic action in a general $P(\phi,X)$ theory. 
As a result, their normalization and dispersion relation differ from those of canonical relativistic particles, and this difference feeds directly into both the decay amplitude and the available phase space.

To make this connection explicit, we adopt a well-motivated k-essence framework in which the late-time dynamics are characterized by a constant normalization $\kappa\equiv P_X$. This assumption is well motivated, since cosmological observations indicate that dark energy closely approximates a slowly evolving component at late times~\cite{Planck2018}, implying that the background field evolution is strongly suppressed~\cite{Caldwell:1997ii,Tsujikawa:2013fta}. In this regime, higher-order time dependence in $P(\phi,X)$ contributes only subleading corrections to the background expansion, while the leading phenomenological effects relevant for perturbations, such as the sound speed and normalization of kinetic fluctuations, are captured by derivatives of $P$ evaluated on the background~\cite{Garriga:1999vw,DeDeo:2003te}. As a result, treating $\kappa$ as constant provides a model-independent parametrization of late-time dark energy that isolates the microphysical properties controlling perturbations without loss of generality for the collider–cosmology connection explored here. In this limit, the sound speed can be written as
\begin{equation}
\Sigma \equiv P_X+2\bar X P_{XX},
\qquad
c_s^2=\frac{P_X}{\Sigma}=\frac{\kappa}{\Sigma},
\label{eq:cs^2_kappaSigma}
\end{equation}
where $\bar X=\frac{1}{2}\dot{\bar\phi}^2$ is the background kinetic energy. This relation can be inverted to express the normalization of fluctuations as
\begin{equation}
\Sigma=\frac{\kappa}{c_s^2},
\end{equation}
which makes clear that the sound speed controls the relative weight of temporal and spatial derivatives in the quadratic action.

It is useful to provide an estimate of the expected magnitude and range of $\kappa$. 
By definition, $\kappa = P_X$ evaluated on the cosmological background controls the normalization of the kinetic term for fluctuations and therefore sets the overall scale relating field gradients to physical energy density. 
In canonical quintessence models, where $P(\phi,X)=X-V(\phi)$, one has $P_X=1$, corresponding to $\kappa=1$ in natural units. 
More generally, in k-essence theories $\kappa$ can deviate from unity, but theoretical consistency and observational constraints impose important bounds on its allowed values.

First, the absence of ghost instabilities requires $P_X>0$, implying $\kappa>0$. 
Second, the requirement that the dark energy equation of state remains close to $w\simeq -1$ at late times implies that the kinetic contribution to the energy density is subdominant, $X P_X \ll V(\phi)$, but does not directly fix the normalization $\kappa$ itself. 
Instead, $\kappa$ determines how fluctuations are canonically normalized relative to the background energy density.

From an effective field theory perspective, it is natural to expect $\kappa$ to be of order unity in the absence of additional symmetries or strong-coupling effects, since it corresponds to the leading coefficient in a derivative expansion. 
Values $\kappa \ll 1$ would indicate a suppressed kinetic term and can lead to strong-coupling behavior for fluctuations, while $\kappa \gg 1$ corresponds to an enhanced kinetic normalization, which can arise in certain non-canonical or screening scenarios.

In the present work, we therefore treat $\kappa$ as a positive parameter of order unity, $\kappa \sim \mathcal{O}(1)$, while allowing for deviations to capture non-canonical dark energy dynamics. 
Importantly, as will be shown in the remainder of this manuscript, decay widths depend on $\kappa$ primarily through inverse powers such as $\kappa^{-1}$ and $\kappa^{-2}$, so that even $\mathcal{O}(1)$ variations can lead to parametrically distinct phenomenological effects.

At the level of the cosmological background, we can use the equation $\rho_\phi + p_\phi = 2X P_X$ to relate the kinetic density $X$ to the dark energy equation of state $w$. This yields
\begin{equation}
X=\frac{\rho_\phi(1+w)}{2\kappa}.
\end{equation}
This expression highlights an important separation of physical roles: the parameter $w$ determines the size of the background kinetic density and therefore controls the kinetic-mixing effect encoded in $Z_a$, while the sound speed $c_s^2$ controls the dynamics of fluctuations and thus governs the production and propagation of dark energy quanta at colliders.

Substituting this result into the expression for the kinetic normalization parameter gives
\begin{equation}
Z_a = 1 - \frac{4XC_8}{\Lambda_8^4}
= 1 - \frac{2C_8\rho_\phi(1+w)}{\kappa \Lambda_8^4},
\end{equation}
demonstrating that $Z_a$ is sensitive to the background evolution but, in this framework, does not depend explicitly on $c_s^2$. 
The sound-speed dependence instead enters through the properties of the produced final-state particles.

We now explicitly derive the dependence of collider observables on the sound speed by taking the standard approach of decomposing the scalar field into a cosmological background and a fluctuation,
\begin{equation}
\phi=\bar\phi(t)+\delta\phi,
\end{equation}
where $\bar\phi(t)$ describes the homogeneous time-dependent background and $\delta\phi$ represents small perturbations around this background, which correspond to the physical quanta that can be produced at colliders.

The dynamics of these fluctuations are obtained by expanding the Lagrangian $P(\phi,X)$ about the background configuration $\rm bg\equiv(\bar\phi(t),\bar X)$~\cite{Garriga:1999vw,ArmendarizPicon:2000dh}. 
We can then write
\begin{equation}
\begin{split}
P(\phi,X)
&=
P(\bar\phi,\bar X)
+ \left.\frac{\partial P}{\partial \phi}\right|_{\rm bg}\delta\phi
+ \left.\frac{\partial P}{\partial X}\right|_{\rm bg}\delta X\\
&+ \frac12 \left.\frac{\partial^2 P}{\partial X^2}\right|_{\rm bg}(\delta X)^2
+\cdots,
\end{split}
\label{eq:P(X)expansion}
\end{equation}
where $\delta X=\dot{\bar\phi}\,\partial_0\delta\phi$ and $\delta X^2=\frac{1}{2}(\partial_0\delta\phi)^2-\frac{1}{2}(\nabla\delta\phi)^2$ contain terms linear and quadratic in $\delta\phi$.  

The zeroth-order term $P(\bar\phi,\bar X)$ contributes only to the background energy density and pressure, and therefore does not affect the dynamics of fluctuations. 
The linear terms vanish upon using the background equations of motion, which ensure that $\bar\phi(t)$ extremizes the action~\cite{Mukhanov:1990me,Weinberg:2008zzc}. 
In this work, we take the quadratic terms in the expansion to be the leading contribution governing the propagation of fluctuations. We also restrict our focus to purely kinetic k-essence models, which comprise a well-established subset of models wherein $P(\phi, X)=P(X)$ explicitly or higher-order $\phi$-derivatives of $P$ are taken to be negligible~\cite{Scherrer:2004kinetickess, Cordero:2016kink-esstop}. In the context of our work, this choice is consistent with the approximate shift-symmetry of $\phi$.

Keeping only terms quadratic in $\delta\phi$ and its derivatives, one obtains the effective k-essence quadratic Lagrangian for the fluctuations~\cite{Garriga:1999vw}. 
In the preferred cosmological frame, this takes the schematic form
\begin{equation}
\Lag^{(2)} \supset \frac12\,\Sigma\,(\partial_0\delta\phi)^2
-\frac12\,\kappa\,(\nabla\delta\phi)^2,
\end{equation}
where $\Sigma$ and $\kappa$ are evaluated on the background. 

We see that time and spatial derivatives enter with different coefficients, reflecting the fact that the time-dependent background $\bar\phi(t)$ spontaneously breaks Lorentz invariance by selecting a preferred rest frame. 
The equation of motion derived from this quadratic action then leads to the dispersion relation
\begin{equation}
\omega^2 = c_s^2\,k^2,
\end{equation}
where the sound speed is given by Eq.~(\ref{eq:cs^2_kappaSigma}).
Thus, the sound speed emerges as the ratio of the coefficients of spatial and temporal kinetic terms, and directly controls the propagation of the dark energy fluctuations that appear as final-state particles in collider processes. 

In order to compute decay rates using standard S-matrix methods, we introduce a canonically normalized field,
\begin{equation}
\phi_c \equiv \sqrt{\Sigma}\,\delta\phi = \sqrt{\frac{\kappa}{c_s^2}}\,\delta\phi,
\end{equation}
so that the fluctuations have a standard kinetic term in time. 
This redefinition implies that derivatives of $\delta\phi$ are related to those of $\phi_c$ by
\begin{equation}
(\partial\delta\phi)^2 = \frac{c_s^2}{\kappa} (\partial\phi_{c})^2,
\end{equation}
which will play a crucial role in determining the effective interaction strength.

Substituting this relation into the derivative cubic interaction and expressing the mediator in terms of its canonically normalized field $a_c$, we obtain an effective interaction of the form
\begin{equation}
\Lag \supset g\, a_c (\partial\phi_{c})^2,
\qquad
g = \frac{C_5}{\Lambda_5}\,\frac{c_s^2}{\kappa}\,\frac{1}{\sqrt{Z_a}}.
\label{eq:ac_partialphic2}
\end{equation}
Thus, we find that the sound speed enters the interaction strength through the canonical normalization of the dark energy fluctuations.

We now compute the decay amplitude for $a_c \to \phi_{c}\phi_{c}$.
In momentum space, the derivative structure in Eq.~\ref{eq:ac_partialphic2} implies that the matrix element is proportional to the Lorentz-invariant product of the outgoing momenta, such that 
$\mathcal{M} = i g\, (p_1 \cdot p_2)$, where $p_1$ and $p_2$ are the momenta of the final-state particles $\phi_c$ and $p_1\cdot p_2 =\omega_1\omega_2-\bm{p}_1\cdot\bm{p}_2$. 
In the rest frame of the mediator $a_c$, energy conservation together with the modified dispersion relation $\omega = c_s |\bm{k}|$ fixes the kinematics of the final state. 
In particular, one finds that the magnitude of the three-momentum scales as $|\bm{p}| \sim m/(2c_s)$, which differs from the case of canonical quintessence by a factor of $1/c_s$. 
Evaluating the invariant product then yields
\begin{equation}
p_1 \cdot p_2 = \frac{m^2}{4}\left(1+\frac{1}{c_s^2}\right),
\end{equation}
and therefore the squared amplitude becomes
\begin{equation}
|\mathcal{M}|^2 = g^2 \frac{m^4}{16}\left(1+\frac{1}{c_s^2}\right)^2.
\end{equation}

We note that the above procedure assumes the cosmological background $(\bar\phi,\bar X)$ is approximately constant over the microscopic spacetime region of the interaction and that the mediator $a_c$ is not boosted with respect to the preferred rest frame selected by this background. The first assumption is reasonable given the fact that the Solar System's speed with respect to the CMB is of order $\beta_{\rm lab}\sim10^{-3}$. For the second assumption, if instead $a_c$ was moving with speed $\beta_{\rm rel}$ with respect to the preferred cosmological frame, a dependence on the quantity $\boldsymbol{\beta_{\rm rel}}\cdot\textbf{k}$ would arise at leading order. This contribution becomes negligible in the case that $\beta_{\rm rel}\ll 1$, which can reasonably occur when $a_c$ is produced near threshold as in the dominant production mechanism of the SM Higgs boson~\cite{PDG:2026review}.

Returning to the decay width calculation, the second ingredient is the two-body phase space, which determines the density of accessible final states. 
In the standard case of canonical quintessence, this quantity depends only on the masses of the particles. 
However, in the present setup the dispersion relation of the final-state quanta is modified, and this leads to a nontrivial dependence of the phase-space measure on the sound speed $c_s$. The starting point is the standard S-matrix phase-space element, constructed from canonically normalized asymptotic states,
\begin{equation}
\dd\Phi_2 = \int \frac{\dd^3 \bm{p}_1}{(2\pi)^3\,2\omega_1}
\frac{\dd^3 \bm{p}_2}{(2\pi)^3\,2\omega_2}
(2\pi)^4 \delta^{(4)}(p - p_1 - p_2),
\end{equation}
where the energies $\omega_i$ now satisfy the modified dispersion relation $\omega_i = c_s |\bm{p}_i|$. Working in the rest frame of the decaying mediator $a_c$, the spatial part of the delta function enforces $\bm{p}_2 = -\bm{p}_1$, while the energy delta function imposes $m = \omega_1 + \omega_2 = 2 c_s |\bm{p}|$, which fixes the magnitude of the three-momentum to be $|\bm{p}| = \frac{m}{2c_s}$ as mentioned previously.

This measure follows from treating the dark-energy quanta as canonically normalized states, not from quantization with respect to an emergent acoustic metric~\cite{Sawicki:2024ryt}, for which the inner product, state normalization, and invariant phase-space measure would instead be defined relative to that effective geometry. In the present work we adopt the collider EFT description, where the states are canonically normalized particles propagating in Minkowski spacetime with the modified dispersion relation $\omega_i = c_s|\bm{p}_i|$. Under these assumptions the usual S-matrix phase-space measure remains valid, with the sound-speed dependence entering through the dispersion relation and the canonical normalization of the dark-energy fluctuations (see section 6 for the relation between the $S$-matrix and acoustic geometry formulations).

Relative to the standard case, this relation shows that a given energy corresponds to a larger spatial momentum when $c_s<1$. 
This modifies both the momentum measure $\dd^3\bm{p}$ and the Jacobian associated with the energy-conserving delta function. 
Performing the angular integrals and evaluating the remaining delta function, one finds that the phase-space measure acquires an overall enhancement factor that depends on the sound speed. 
The final result can be written as
\begin{equation}
\int \dd\Phi_2 = \frac{1}{8\pi c_s^3}.
\end{equation}

The factor of $c_s^{-3}$ has a clear physical interpretation: for subluminal propagation ($c_s<1$), the same total energy can be distributed among final states with larger momenta, effectively increasing the density of accessible states. 
This enhancement plays a crucial role in amplifying the decay width into dark energy quanta and is one of the key mechanisms through which the sound speed enters collider observables. In the canonical quintessence limit $c_s=1$, this expression reduces to the standard massless two-body result,
\begin{equation}
\int \dd\Phi_2\Big|_{c_s=1}=\frac{1}{8\pi},
\end{equation}
so the factor $c_s^{-3}$ directly quantifies the departure from the ordinary relativistic phase space.

Combining the amplitude and phase space, the decay width is found to be
\begin{equation}
\Gamma(a_c \to \phi_{c}\phi_{c})
=
\frac{g^2 m^3}{512\pi}\,
\frac{(1+c_s^2)^2}{c_s^3}.
\end{equation}
Substituting the expression for the effective coupling $g$ and expressing the result in terms of the original Lagrangian parameters yields the central result
\begin{equation}
\Gamma(a_c \to \phi_c\phi_c)
=
\frac{C_5^2}{512\pi}\,
\frac{m_a^3}{\kappa^2 \Lambda_5^2}\,
Z_a^{-5/2}\,
\frac{(1+c_s^2)^2}{c_s^3}.
\label{eq:Gamma_phiphi_maZa_k}
\end{equation}

This expression makes explicit that the sound speed controls the invisible decay width through the factor
\begin{equation}
\mathcal{G}(c_s) = \frac{(1+c_s^2)^2}{c_s^3},
\end{equation}
which grows rapidly for $c_s^2 \ll 1$. 
Physically, this enhancement arises from two effects acting in concert: the derivative structure of the interaction, which introduces additional momentum dependence, and the modified phase space, which increases the density of accessible final states.

The collider implications of this result are immediate. 
Since the invisible width enters directly into the total width 
$\Gamma_{\rm tot} \simeq \Gamma(a\to t\bar t) + \Gamma(a\to \chi\bar\chi) + \Gamma(a\to \phi\phi)$,
the sound speed controls the resonance lineshape and branching ratios. 
Decreasing $c_s^2$ leads to an enhanced invisible width, which in turn broadens the resonance and suppresses visible branching fractions. 
These effects persist across the full mediator mass range, and are not limited to the near-threshold region.

\begin{figure}[ht]
    \centering
    \includegraphics[width=\textwidth]{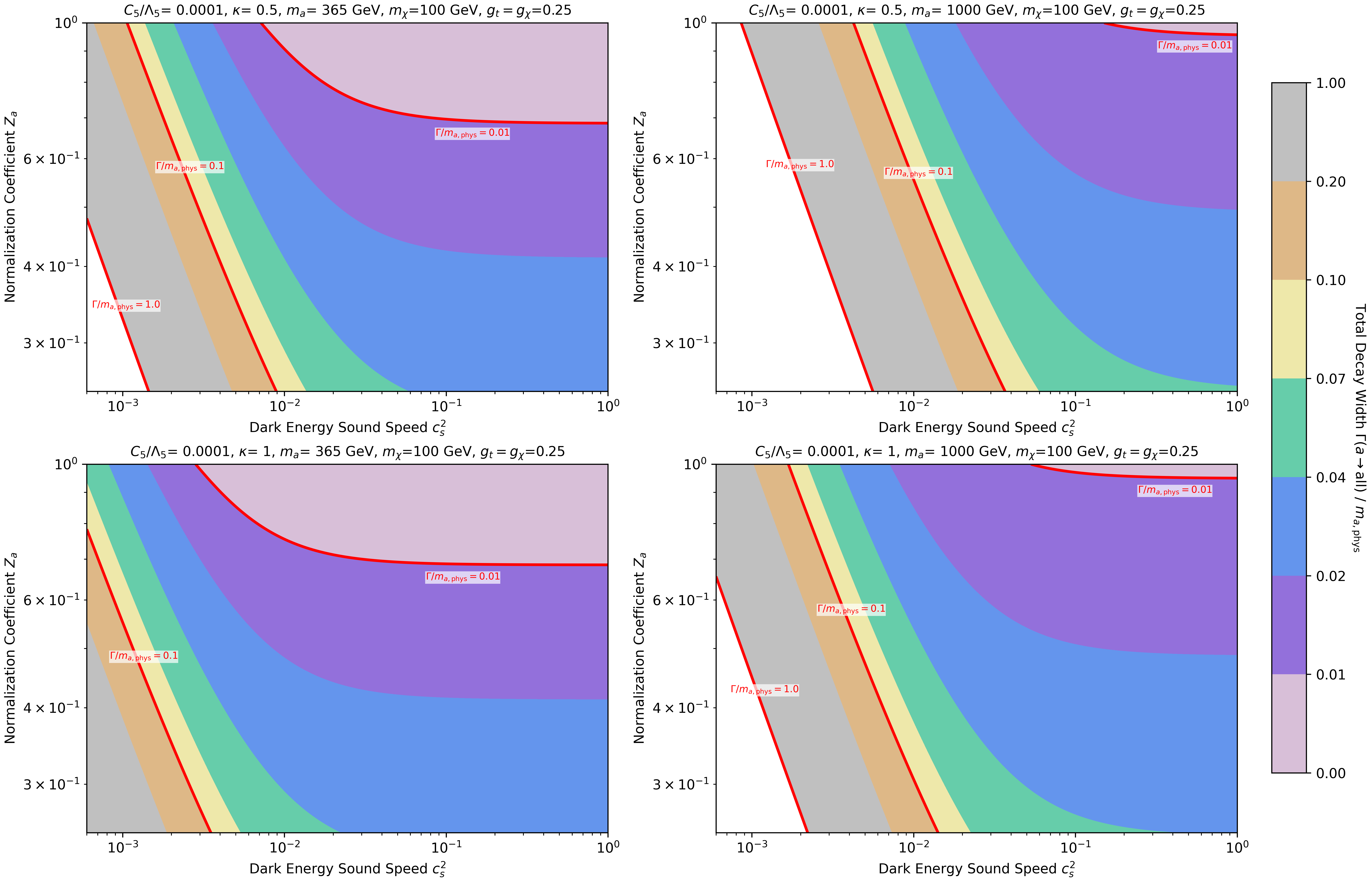}
    \caption{Color maps of the total normalized decay width $\Gamma_{\rm tot}/m_{a,\rm phys}$, with discrete colors corresponding to different intervals between a normalized width of $0$ and $1$. We plot $c_s^2$ on the horizontal axis and $Z_a$ on the vertical axis, and vary the parameters $m_a$ and $\kappa$ across the four plots. Contours for $\Gamma_{\rm tot}/m_{a,\rm phys}=0.01$, $0.1$, and $1$ are drawn in red.}
    \label{fig:totwidth_cs2Za}
\end{figure}
To illustrate these features, Fig.~(\ref{fig:totwidth_cs2Za}) shows color maps of the normalized total width, $\Gamma_{\rm tot}/m_{a,\rm phys}$, as a function of $c_s^2$ and $Z_a$. The four panels correspond to two representative mediator masses, $m_a=365$ and $1000~\mathrm{GeV}$, chosen to probe the region near the $t\bar{t}$ production threshold and a substantially heavier resonance, respectively, and two values of $\kappa$, namely $\kappa=0.5$ and $1$. The remaining parameters are fixed to $m_\chi=100~\mathrm{GeV}$ and $g_t=g_\chi=0.25$, as indicated in the figure titles. These values are typical benchmarks used in dark sector collider searches~\cite{CMS:2024zqs}. The discrete color scale, shown on the right, spans values of $\Gamma_{\rm tot}/m_{a,\rm phys}$ between 0 and 1.
We focus primarily on the region $0.01 \lesssim \Gamma_{\rm tot}/m_{a,\rm phys} \lesssim 0.2$ since widths at the percent level are characteristic of many SM resonances and are also consistent with the narrow-width interpretation of the recently reported di-top excesses by CMS and ATLAS~\cite{CMS:2025kzt,ATLAS:2026dbe}. At the same time, broader resonances are well motivated in many BSM extensions, making it important to understand how the width depends on the underlying dark energy microphysics. The resonance width also plays a central role in collider phenomenology, influencing both signal modeling and the validity of common theoretical approximations~\cite{Shen:2025nkr}. As shown in Fig.~(\ref{fig:totwidth_cs2Za}), the normalized width generally increases as either $c_s^2$ or $Z_a$ decreases. The relative importance of these parameters, however, varies across the parameter space. For $c_s^2$ close to unity, the width is primarily controlled by $Z_a$, while the dependence on $c_s^2$ becomes increasingly pronounced as one approaches the $\Gamma_{\rm tot}/m_{a,\rm phys}\approx0.1$ contour (shown in red). Below this threshold, the width grows rapidly with decreasing $c_s^2$, eventually reaching the perturbative unitarity limit $\Gamma_{\rm tot}/m_{a,\rm phys}=1$.
These results suggest that collider measurements of the mediator width could provide a direct probe of the dark energy sector. In particular, the width is highly sensitive to $c_s^2$ when $c_s^2 \ll 1$, while retaining significant sensitivity to $Z_a$ throughout the parameter space considered. Consequently, precision measurements of the resonance line shape and decay width at colliders could offer a unique window into the microphysical properties of the dark energy field $\phi$, complementing traditional cosmological probes. 

\begin{figure}[ht]
    \centering
    \includegraphics[width=\textwidth]{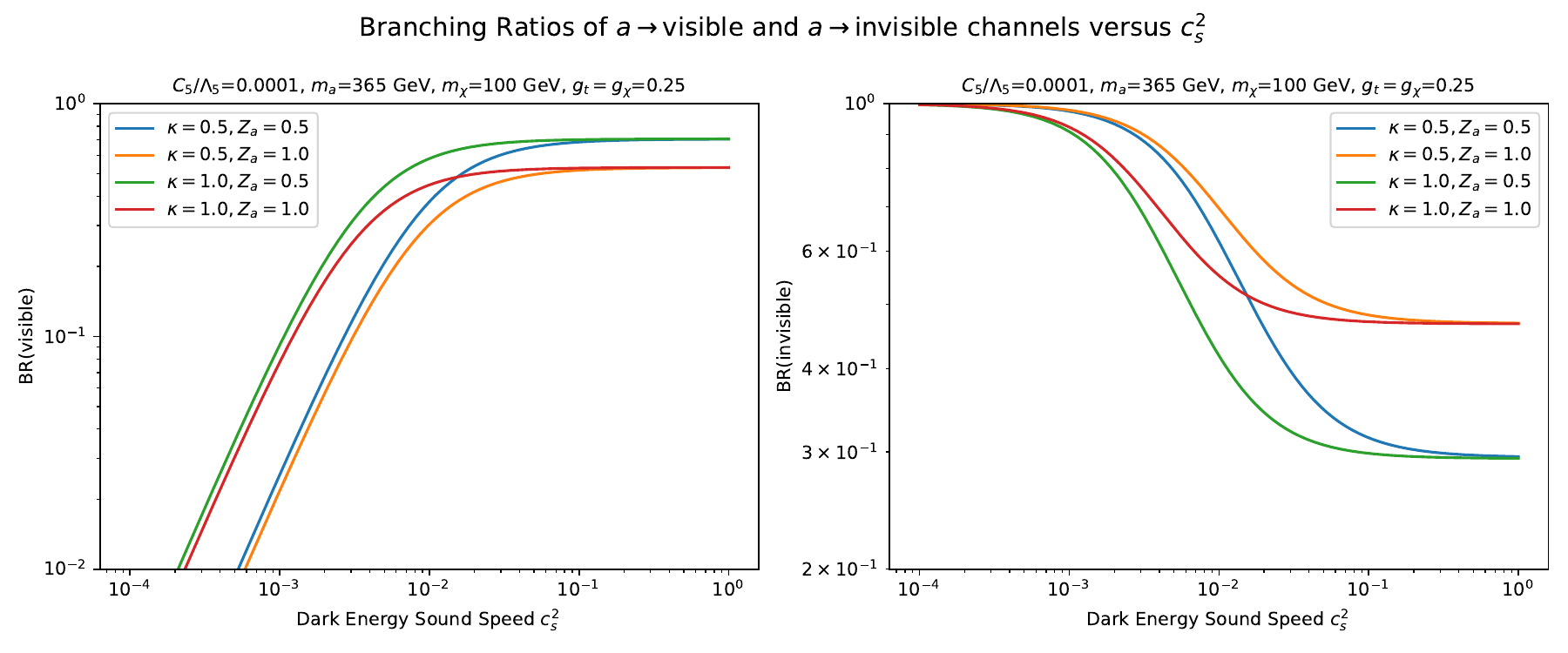}
    \caption{Branching ratios of the visible and invisible decay channels of $a$, defined by Eq.~(\ref{eq:brvisible}) and Eq.~(\ref{eq:brinvisible}), versus $c_s^2$. The four curves in each of the three plots vary the parameters $\kappa$ and $Z_a$ as shown in the legend.}
    \label{fig:branchingratios_cs2}    
\end{figure}

Next, we plot the branching ratios of the visible and invisible parts of the decay versus $c_s^2$ in Fig.~(\ref{fig:branchingratios_cs2}). Here, we define
\begin{equation}
    \textrm{BR}(\textrm{visible})=\textrm{BR}(a\to t\bar t)=\frac{\Gamma(a\to t\bar t)}{\Gamma_{\rm tot}}
\label{eq:brvisible}
\end{equation}
and
\begin{equation}
\begin{split}
    \textrm{BR}(\textrm{invisible})&=\textrm{BR}(a\to\chi\bar\chi)+\textrm{BR}(a\to\phi\phi)\\
    &=\frac{\Gamma(a\to \chi\bar\chi)+\Gamma(a\to\phi\phi)}{\Gamma_{\rm tot}}.
\end{split}
\label{eq:brinvisible}
\end{equation}
The four curves correspond to different combinations of $(\kappa, Z_a)$, with other parameters set to $C_5/\Lambda_5=10^{-4}\ \text{GeV}^{-1}$, $m_\chi = 100$ GeV, and $g_{t}=g_{\chi}=0.25$. This plot displays that the $a\to\rm invisible$ channel behaves quite differently from the $a\to \rm visible$ channel, which is due to the derivative coupling $\sim a(\partial\phi)^2$ in the Lagrangian term coupling $a$ to the dark energy field $\phi$. We again observe that $Z_a$ dominates the behavior for $c_s^2\sim 1$. Meanwhile, for lower sound speeds we see that $\kappa$ significantly affects the branching ratios, with an increase in $\kappa$ generally causing an increase in $\textrm{BR}(\textrm{visible})$ and a decrease in $\textrm{BR}(\textrm{invisible})$. For all parameter values, including the canonical case of $(\kappa,Z_a)=(1,1)$, there is a clear dependency on $c_s^2$ in both channels with high sensitivity at lower sound speeds. 

\begin{figure}[ht]
    \centering
    \includegraphics[width=.8\linewidth, trim=2cm 0 9cm 1cm, clip]{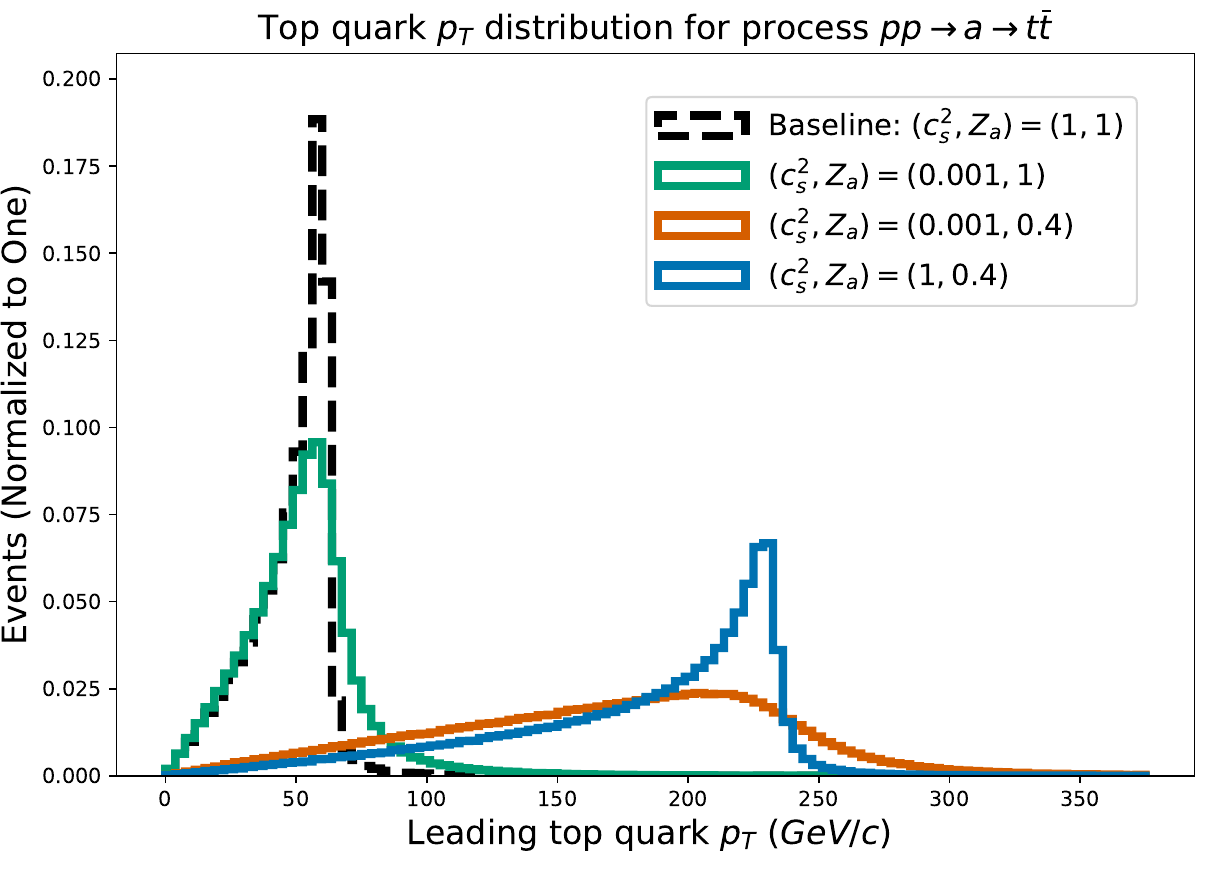}
    \caption{Histograms of the transverse momentum of a final state top quark for the process indicated, with events normalized to one. The four curves correspond to different combinations of $(c_s^2,Z_a)$, with the curve in black serving as the baseline scenario with $C_5/\Lambda_5=0\ \text{GeV}^{-1}$.}
    \label{fig:topptdist}
\end{figure}

Finally, we investigate the impact of dark energy microphysics on collider kinematics by studying the transverse-momentum $(p_T)$ distribution of the top quarks in the process $pp \to a \to t\bar{t}$, shown in Fig.~\ref{fig:topptdist}. The distributions are generated using the matrix-element event generator \texttt{MadGraph5\_aMC@NLO} \cite{Alwall:2014madgraph5}, assuming proton-proton collisions at $\sqrt{s}=13.6$ TeV, and analyzed with \texttt{MadAnalysis5} \cite{Conte:2013madanalysis5}. At leading order, the top and antitop quarks are produced back-to-back in the transverse plane and therefore share identical $p_T$ distributions.
The four histograms correspond to different choices of the dark energy parameters $(c_s^2, Z_a)$, as indicated in the legend. All other model parameters are fixed to the values used in the lower-left panel of Fig.~\ref{fig:totwidth_cs2Za}. For the black histogram, we additionally set $C_5/\Lambda_5=0\ \text{GeV}^{-1}$, corresponding to a canonically normalized mediator completely decoupled from the dark energy field. It serves as a ``baseline" reference against which the effects of nontrivial dark energy dynamics can be assessed.
Comparing the black and blue histograms, or equivalently the green and orange histograms, demonstrates that reducing $Z_a$ from unity shifts the distribution toward larger $p_T$ values while also introducing some broadening. This behavior reflects the increase in the physical mediator mass predicted by Eq.~(\ref{eq:ma_phys}). In contrast, decreasing $c_s^2$ primarily broadens the distribution without significantly shifting its peak position, as can be seen by comparing the black (baseline) and green histograms or the blue and orange histograms. The orange distribution, corresponding to simultaneously reduced $c_s^2$ and $Z_a$, exhibits the most pronounced broadening. 
From an experimental perspective, these results suggest that dark energy microphysics can leave observable imprints on collider kinematic distributions. In particular, sufficiently small values of $c_s^2$ can generate resonance shapes that depart noticeably from the narrow-width approximation and standard Breit--Wigner expectations. Although the present study is based on a simplified parton-level analysis that neglects effects such as initial-state radiation, top-quark decays, parton showering, and detector reconstruction, the qualitative trends are clear: variations in either the normalization parameter $Z_a$ or the sound speed $c_s^2$ can produce substantial distortions in the observed top-quark $p_T$ spectrum. These findings highlight the potential of precision collider measurements not only to discover new mediators, but also to probe the propagation properties and underlying microphysics of the dark energy field itself.

Taken together, these results demonstrate that collider measurements of mediator widths and kinematics can probe the propagation properties of dark energy fluctuations. 
In particular, the dependence on $c_s^2$ provides a direct window into the microphysics of dark energy, establishing a novel and quantitative connection between high-energy collider experiments and cosmological dynamics.

\subsection{Validity of the P(X) expansion and linear dispersion relation}

The calculations presented in the previous section are based on an expansion of the dark energy Lagrangian $P(X)$ about the late-time cosmological background, retaining the quadratic terms in the fluctuations. The resulting expressions therefore describe the leading-order dynamics of the dark-energy fluctuations and give rise to the linear dispersion relation
$\omega_{\mathbf{k}}=c_s\abs{\mathbf{k}}$. We now discuss the regime of validity of the treatment above. There are two conceptually distinct sources of consideration to this description. First, higher orders in the fluctuation expansion of $P(X)$ generate nonlinear interactions among the dark energy excitations and ultimately determine the strong-coupling scale of the theory, which we denote as $\Lambda_{\rm SC}$. Second, higher spatial-derivative operators in the full EFT can modify the quadratic kinetic operator itself and hence the dispersion relation. We denote the scale suppressing such operators by $\Lambda_{\rm EFT}$. Below we will show that both scales are model dependent and cannot be determined uniquely within the model-independent $P(X)$ parametrization considered here.

We begin by studying the strong-coupling scale $\Lambda_{\rm SC}$, following the expansion approach illustrated in Eq.~(\ref{eq:P(X)expansion}) but expanding up to terms cubic in $\delta\phi$ and $\delta X+\delta^2 X$ and keeping terms up to $\mathcal{O}(\delta\phi^3)$. This process yields the cubic Lagrangian
\begin{equation}
    \Lag^{(3)}=P_{XX}\,\delta X\,\delta^2X+\frac{1}{6}P_{XXX}\,(\delta X)^3.
\end{equation}
We can then plug in expressions for $\delta X,\,\delta^2 X$ in terms of derivatives of $\delta\phi$ and write the cubic Lagrangian in terms of the canonically normalized field $\phi_c=\sqrt{\Sigma}\,\delta\phi$. Moreover, we use the relationships $\Sigma=P_X+2\bar XP_{XX}$ and $\Sigma=\kappa/c_s^2$ to simplify. This yields 
\begin{equation}
\begin{aligned}
    \Lag^{(3)}&=\left(\frac{c_s(1-c_s^2)}{2\sqrt{2\bar X\kappa}}+\frac{\sqrt{2}\,\bar X^{3/2}P_{XXX}\,c_s^3}{3\,\kappa^{3/2}}\right)\dot{\phi_c}^3-\frac{c_s(1-c_s^2)}{2\sqrt{2\bar X\kappa}}\,\dot{\phi_c}(\nabla\phi_c)^2\\
    &=(g_1+g_2)\,\dot{\phi_c}^3-g_1\,\dot{\phi_c}(\nabla\phi_c)^2,
\end{aligned}
\end{equation}
where we define $g_1\equiv \frac{c_s(1-c_s^2)}{2\sqrt{2\bar X\kappa}}$ and $g_2\equiv \frac{\sqrt{2}\,\bar X^{3/2}P_{XXX}\,c_s^3}{3\,\kappa^{3/2}}$. Note that these terms are cubic in the power of the field, meaning that they represent three-point interactions of dark energy quanta. These cubic interactions, in addition to quartic and higher interactions, set the strong-coupling scale $\Lambda_{\rm SC}$ of the theory. They therefore do not modify the linearized equation of motion or the linear dispersion relation. Instead, they characterize the leading nonlinear corrections to the quadratic theory.

We can investigate the behavior of the cubic terms in more detail by defining energy scales $\Lambda_{\rm SC,1}$ and $\Lambda_{\rm SC,2}$ such that $g_1=1/\Lambda_{\rm SC,1}^2$ and $g_2=1/\Lambda_{\rm SC,2}^2$ (since $\text{dim}(g_1)=\text{dim}(g_2)=-2$ in energy units). In order for these three-point interactions contributing to strong-coupling to be suppressed, $\Lambda_{\rm SC,1}$ and $\Lambda_{\rm SC,2}$ must be larger than the characteristic energy scale of a dark energy quanta in the collider. Solving for both energy scales explicitly gives 
\begin{equation}
\begin{aligned}
    &\Lambda_{\rm SC,1}=\left(\frac{2\sqrt{2\bar X\kappa}}{c_s(1-c_s^2)}\right)^{1/2},\\
    &\Lambda_{\rm SC,2}=\left(\frac{3\,\kappa^{3/2}}{\sqrt{2}\,\bar X^{3/2}P_{XXX}\,c_s^3}\right)^{1/2}.
\end{aligned}    
\end{equation}
The above form allows us to see that both scales approach infinity in the quintessence limit $c_s^2,\,\kappa\to1$, as expected, and in the general k-essence setting vary inversely with $c_s$. Despite these observations, both $\Lambda_{\rm SC,1}$ and $\Lambda_{\rm SC,2}$ depend on $\kappa$ and $\bar X$ (and higher derivatives of $P$), which are model-dependent quantities and thus treated as phenomenological parameters in our model-independent approach. Consequently, the sound speed alone does not determine the strong-coupling scale of a general k-essence theory. We therefore do not apply any explicit constraints but assume that they behave such that these energy scales, and energy scales associated with higher-order interactions of dark energy quanta, are sufficiently high.

We next consider the separate question of the validity of the linear dispersion relation. As stated above, the higher-order terms in the $P(X)$ expansion correspond to interactions between fluctuations, and thus do not contribute to the linear dispersion relation. In fact, in order for a term in the Lagrangian to contribute to the dispersion relation, it must be quadratic in the fields. For example, a higher-derivative operator of the schematic form $\frac{\alpha}{\Lambda_{\mathrm{EFT}}^2}
\left(\nabla^2 \delta\phi\right)^2$ would generate a correction of the form $\omega_{\mathbf{k}}^2
= c_s^2|\mathbf{k}|^2
+
\widetilde{\alpha}
\frac{|\mathbf{k}|^4}{\Lambda_{\rm EFT}^2}
+\cdots$, where $\widetilde{\alpha}$ includes the appropriate background normalization and Wilson coefficient. The precise coefficients and the value of $\Lambda_{\rm EFT}$ depend on the UV completion and cannot be inferred from $P(X)$ alone. We therefore assume an EFT cutoff scale such that $k=\frac{m}{2c_s}<\Lambda_{\rm EFT}$ for the parameter space of interest. This ensures the validity of Eq.~(\ref{eq:fullLag}) and allows for the use of our linear dispersion relationship at collider scale. 

The complete validity requirements relevant for the leading treatment employed in this work may therefore be summarized schematically as $E_\phi
<
\Lambda_{\rm SC}$ and 
$\frac{m}{2c_s} <
\Lambda_{\rm EFT}$. Here $\Lambda_{\rm SC}$ denotes the lowest relevant nonlinear interaction or perturbative-unitarity scale of the specified $P(X)$ model. The first condition ensures perturbative control over the nonlinear interactions of the dark-energy fluctuations, while the second ensures that higher-derivative corrections to the quadratic kinetic operator remain suppressed. Since both scales depend on physics beyond the model-independent quantities $c_s$ and $\kappa$, our collider predictions should be interpreted as applying to the set of k-essence models for which these inequalities are satisfied.

\section{Relation Between the Collider $S$-matrix \& Acoustic-Geometry Descriptions}
\label{sec}
In the preceding discussion, the dark-energy fluctuations are treated as canonically normalized states with the modified dispersion relation $\omega_{\mathbf{k}}=c_s|\mathbf{k}|$, and the corresponding decay rates are evaluated using an $S$-matrix description in the preferred cosmological frame. Since perturbations of a $P(\phi,X)$ theory may alternatively be formulated as fields propagating in an effective acoustic geometry~\cite{Sawicki:2024ryt}, it is useful to clarify the relation between these two descriptions. The essential assumption underlying the collider calculation is that the late-time dark-energy background is homogeneous and varies negligibly over the microscopic spacetime region associated with the collider hard process. In this local limit, $\dot{\bar\phi}$ and the background quantities $\Sigma$ and $\kappa$ may be treated as constants. Time and spatial translations therefore remain good symmetries on the scale of the collider event, implying conserved energy and momentum and permitting the usual construction of asymptotic momentum eigenstates and a local $S$-matrix. 
In this section we provide the calculations which show that for the homogeneous late-time background and separation of scales assumed in this work, the collider $S$-matrix calculation is the locally constant-background limit of the acoustic-geometry description. The two formulations yield the same characteristic propagation, the same canonically normalized one-particle states, and hence the same
leading decay rate.

\subsection{Canonical quantization in the local collider EFT}
\label{subsec}
First, we discuss canonical quantization in the local collider EFT and justify the use of the phase-space measure adopted in the previous sections. For a homogeneous background, the quadratic action for the dark-energy fluctuation $\delta\phi$ is $S_{\phi}^{(2)} =
\frac{1}{2}
\int d^4x
\left[
\Sigma
\left(\partial_0\delta\phi\right)^2
-
\kappa
\left(\boldsymbol{\nabla}\delta\phi\right)^2
\right]$.
Introducing the canonically normalized fluctuation $\phi_c
\equiv
\sqrt{\Sigma}\delta\phi$, and treating $\Sigma$ and $\kappa$ as constant over the hard process gives
\begin{equation}
S_{\phi_c}^{(2)}
=
\frac{1}{2}
\int d^4x
\left[
\left(\partial_0\phi_c\right)^2
-
c_s^2
\left(\boldsymbol{\nabla}\phi_c\right)^2
\right].
\label{eq:action}
\end{equation}
Canonical normalization is therefore fixed by the quadratic action rather than being an arbitrary convention. In particular, the conjugate momentum is $\Pi_c
=
\frac{\partial\mathcal{L}^{(2)}}
{\partial(\partial_0\phi_c)}
=
\partial_0\phi_c$, and canonical quantization imposes $\left[
\phi_c(t,\mathbf{x}),
\Pi_c(t,\mathbf{y})
\right]
=
i\delta^{(3)}(\mathbf{x}-\mathbf{y})$. The equation of motion following from Eq.~\eqref{eq:action} is
\begin{equation}
\ddot{\phi}_c
-
c_s^2\nabla^2\phi_c
=
0,
\end{equation}
so a plane-wave mode $\phi_c
\propto
e^{-i\omega t+i\mathbf{k}\cdot\mathbf{x}}$ satisfies $\omega_{\mathbf{k}}^2
=
c_s^2|\mathbf{k}|^2$. Therefore, the canonically normalized field has the mode expansion
\begin{equation}
\phi_c(x)
=
\int
\frac{d^3k}{(2\pi)^3}
\frac{1}{\sqrt{2\omega_{\mathbf{k}}}}
\left[
a_{\mathbf{k}}
e^{-i\omega_{\mathbf{k}}t+i\mathbf{k}\cdot\mathbf{x}}
+
a_{\mathbf{k}}^\dagger
e^{+i\omega_{\mathbf{k}}t-i\mathbf{k}\cdot\mathbf{x}}
\right],
\label{eq}
\end{equation}
with $[a_{\mathbf{k}},a_{\mathbf{q}}^\dagger]
=
(2\pi)^3
\delta^{(3)}(\mathbf{k}-\mathbf{q})$. The factor $1/\sqrt{2\omega_{\mathbf{k}}}$ can equivalently be read from the pole structure of the canonical propagator, $D_F(\omega,\mathbf{k})
= i (\omega^2-c_s^2|\mathbf{k}|^2+i\epsilon)^{-1}$. Near the positive-frequency pole, $D_F(\omega,\mathbf{k})
\simeq i (2\omega_{\mathbf{k}})^{-1} (\omega-\omega_{\mathbf{k}}+i\epsilon)^{-1}$, so the external-state normalization gives the phase-space measure contributes the usual factor $\frac{1}{2\omega_{\mathbf{k}}}$. Consequently, despite the modified dispersion relation $\omega_{\mathbf{k}}=c_s|\mathbf{k}|$, the phase-space measure for a canonically normalized dark-energy quasiparticle is $\frac{d^3k}{(2\pi)^3\,2\omega_{\mathbf{k}}}$.  In particular, the two-body phase space for the decay
$a\rightarrow\phi_c\phi_c$ is $d\Phi_2 = \frac{d^3k_1}{(2\pi)^3\,2\omega_1}\frac{d^3k_2}{(2\pi)^3\,2\omega_2}(2\pi)^4 \delta^{(4)}(p_a-k_1-k_2)$. This result follows from canonical quantization and the propagator residue and does not require boost invariance.
\subsection{Equivalent acoustic-geometry formulation}
\label{subsec}
Next, we show that the acoustic geometry and modified dispersion relation used in our work describe the same characteristic propagation. Equivalently, for constant $\Sigma$ and $\kappa$, the acoustic geometry provides an alternative representation of the quadratic Lagrangian operator. We first define the effective acoustic metric
\begin{equation}
G_{\mu\nu} = \begin{pmatrix}
        -1 & 0 \\
        0 & c_s^{-2}\,\mathbf{1}_{3}
\end{pmatrix}
,
\qquad
G^{\mu\nu} =
\begin{pmatrix}
        -1 & 0 \\
        0 & c_s^2\,\mathbf{1}_3
\end{pmatrix},
\label{eq}
\end{equation}
for which the corresponding acoustic line element is $ds_{\rm ac}^2 = G_{\mu\nu}dx^\mu dx^\nu = -dt^2 + \frac{d\mathbf{x}^2}{c_s^2}$ and the determinant of the acoustic metric is $G \equiv \det(G_{\mu\nu}) = -c_s^{-6}$, so that $\sqrt{-G} = c_s^{-3}$.  Introducing $\mathcal{A}
\equiv
\kappa c_s$, one finds
$\mathcal{A}\sqrt{-G}
=
\kappa c_s^{-2}
=
\Sigma$. The original quadratic action can therefore be written exactly in the manifestly acoustic-geometric form
\begin{equation}
S_{\phi}^{(2)} = -\frac{1}{2} \int d^4x\, \mathcal{A}\sqrt{-G}\, G^{\mu\nu} \partial_\mu\delta\phi\, \partial_\nu\delta\phi
\end{equation}
with a temporal contribution of $-\frac{1}{2}
\mathcal{A}\sqrt{-G}
G^{00}
(\partial_0\delta\phi)^2
=
\frac{1}{2}
\Sigma
(\partial_0\delta\phi)^2$, and a spatial contribution of 
$-\frac{1}{2}
\mathcal{A}\sqrt{-G}
G^{ij}
\partial_i\delta\phi
\partial_j\delta\phi
=
-\frac{1}{2}
\kappa
(\boldsymbol{\nabla}\delta\phi)^2$. Thus, for constant $\Sigma$ and $\kappa$, 
\begin{equation}
S_{\phi,\mathrm{ac}}^{(2)}
=
S_{\phi,\mathrm{EFT}}^{(2)}.
\end{equation}
The acoustic metric is therefore an alternative geometric representation of the same quadratic fluctuation operator used in the collider EFT. Varying the acoustic-geometric action with respect to the dark-energy fluctuation $\delta\phi$ gives
\begin{equation}
\frac{1}
{\mathcal{A}\sqrt{-G}}
\partial_\mu
\left[
\mathcal{A}\sqrt{-G}
G^{\mu\nu}
\partial_\nu\delta\phi
\right]
=
0.
\label{eq}
\end{equation}
For the homogeneous background considered here, this reduces to $G^{\mu\nu}
\partial_\mu\partial_\nu\delta\phi
=
0$, or explicitly $-\partial_t^2\delta\phi
+
c_s^2\nabla^2\delta\phi
=
0$. Therefore, the acoustic description gives the same characteristic relation,
$\omega^2 = c_s^2|\mathbf{k}|^2$, as the canonical collider EFT.

\subsection{Equivalence of the one-particle normalization}
The equivalence between the acoustic-geometry and canonical EFT
descriptions can be seen particularly clearly at the level of the one-particle normalization. We start with the Klein--Gordon inner product associated
with the acoustic action: 
\begin{equation}
(u,v)_{\rm ac} = i \mathcal{A} \int_{\Sigma_t} d\Sigma^\mu \left( u^* \partial_\mu v - v \partial_\mu u^* \right).
\end{equation}
For a constant-$t$ hypersurface, the induced spatial metric is $h_{ij} = c_s^{-2}\delta_{ij}$ such that 
$\sqrt{h} = c_s^{-3}$, while the future-directed unit normal is  $n^\mu=(1,0,0,0)$. Hence $d\Sigma^\mu = n^\mu\sqrt{h}\,d^3x = c_s^{-3}d^3x\,(1,0,0,0)$ and the Klein--Gordon inner product associated with the acoustic action gives
\begin{align}
(u,v)_{\rm ac} = i A c_s^{-3} \int d^3x \left( u^*\dot{v} - v\dot{u}^* \right) = i\Sigma \int d^3x \left( u^*\dot{v} - v\dot{u}^* \right).
\end{align}
This is precisely the symplectic product obtained directly from the original quadratic EFT. Thus, the acoustic and collider-EFT formulations possess share the same phase-space dynamics, and consequently dictate an identical normalization for their Hilbert spaces. To make this equivalence explicit, consider the positive-frequency mode functions $u_{\mathbf{k}}(t,\mathbf{x})
    =
    \frac{1}{\sqrt{2\Sigma\omega_{\mathbf{k}}}}\,
    e^{-i\omega_{\mathbf{k}}t
      +i\mathbf{k}\cdot\mathbf{x}},
    \omega_{\mathbf{k}}
    =
c_s|\mathbf{k}|$. Using $(u,v)_{\rm ac} =
i\Sigma \int d^3x \left( u^*\dot{v} - v\dot{u}^* \right)$, their inner product is
\begin{align}
    (u_{\mathbf{k}},u_{\mathbf{q}})_{\rm ac}
    &=
    i\Sigma
    \int d^3x
    \left(
        u_{\mathbf{k}}^*\dot{u}_{\mathbf{q}}
        -
        u_{\mathbf{q}}\dot{u}_{\mathbf{k}}^*
    \right)
    \nonumber\\
    &=
    \Sigma
    (\omega_{\mathbf{k}}+\omega_{\mathbf{q}})
    \frac{1}
    {\sqrt{2\Sigma\omega_{\mathbf{k}}}
     \sqrt{2\Sigma\omega_{\mathbf{q}}}}
    \int d^3x\,
    e^{i(\mathbf{q}-\mathbf{k})\cdot\mathbf{x}}
    \nonumber\\
    &=
    \frac{\omega_{\mathbf{k}}+\omega_{\mathbf{q}}}
    {2\sqrt{\omega_{\mathbf{k}}\omega_{\mathbf{q}}}}
    (2\pi)^3
    \delta^{(3)}(\mathbf{k}-\mathbf{q}).
\end{align} Since the delta function enforces $\mathbf{k}=\mathbf{q}$ and therefore $\omega_{\mathbf{k}}=\omega_{\mathbf{q}}$, this reduces to $(u_{\mathbf{k}},u_{\mathbf{q}})_{\rm ac}
    =
    (2\pi)^3
\delta^{(3)}(\mathbf{k}-\mathbf{q})$. Finally, defining the canonically normalized fluctuation by $\phi_c =
\sqrt{\Sigma}\,\delta\phi$, the corresponding mode functions are $u_{\mathbf{k}}^{(c)}
=
    \sqrt{\Sigma}\,
    u_{\mathbf{k}}
=
    \frac{1}{\sqrt{2\omega_{\mathbf{k}}}}\,
    e^{-i\omega_{\mathbf{k}}t
      +i\mathbf{k}\cdot\mathbf{x}}$.
This is exactly the mode normalization obtained by directly quantizing
the canonically normalized EFT. We therefore find that the acoustic Klein--Gordon normalization is equivalent to the canonical EFT normalization for constant under the assumptions of this work. In particular, the canonically normalized modes carry the usual factor
$1/\sqrt{2\omega_{\mathbf{k}}}$, with
$\omega_{\mathbf{k}}=c_s|\mathbf{k}|$. Consequently, the corresponding
one-particle external-state normalization leads to the phase-space
factor $d^3k/[(2\pi)^3 2\omega_{\mathbf{k}}]$. This result follows from the identical phase-space dynamics of the two descriptions and from the assumption that the cosmological background is homogeneous and stationary over the microscopic spacetime region relevant to the collider process.

\subsection{Validity of the locally constant-background approximation}
\label{subsec}
The equivalence above relies on the separation between the microscopic collider scale and the characteristic scale over which the cosmological background evolves. If $\Sigma$, $\kappa$, or $c_s$ acquire appreciable spacetime dependence over the process ($\Sigma=\Sigma(t,\mathbf{x})$, $\kappa=\kappa(t,\mathbf{x})$, $c_s=c_s(t,\mathbf{x})$ ), 
then $G_{\mu\nu}=G_{\mu\nu}(x)$ is a genuinely spacetime-dependent acoustic geometry. The fluctuation equation remains
$\frac{1}
{\mathcal{A}\sqrt{-G}}
\partial_\mu
\left[
\mathcal{A}\sqrt{-G},
G^{\mu\nu}
\partial_\nu\delta\phi
\right]
=
0$, but where $\mathcal{A}=\mathcal{A}(t,\mathbf{x})$. Equivalently, if the prefactor $\mathcal{A}$ is absorbed into the definition of the effective metric, this may
be expressed in the conventional curved-spacetime form $\frac{1}{\sqrt{-G}}\,
    \partial_\mu
    \left[
        \sqrt{-G}\,
        G^{\mu\nu}
        \partial_\nu\delta\phi
    \right]
    =
0$. Unlike the constant-background case, the coefficients of the wave
operator in $\frac{1}{\mathcal{A}\sqrt{-G}}\,
    \partial_\mu
    \left[
        \mathcal{A}\sqrt{-G}\,
        G^{\mu\nu}
        \partial_\nu\delta\phi
    \right]
    =
0$ and $\frac{1}{\sqrt{-G}}\,
    \partial_\mu
    \left[
        \sqrt{-G}\,
        G^{\mu\nu}
        \partial_\nu\delta\phi
    \right]
    =
0$ are now spacetime dependent. Consequently, the fluctuation modes cannot in general be written globally as plane waves of the form $\delta\phi_{\mathbf{k}}(t,\mathbf{x})
    \propto
e^{-i\omega t+i\mathbf{k}\cdot\mathbf{x}}$ with constant $\omega$ and $\mathbf{k}$. In particular, a time-dependent background generally removes exact time-translation invariance, and a global particle interpretation requires suitable asymptotic or adiabatic regions. This is the regime in which the full spacetime-dependent acoustic-geometry formulation is essential.

For the late-time cosmological background considered here, the accuracy of the local approximation may be estimated parametrically. If the acoustic background varies on a characteristic cosmological scale $L_{\rm bg} \sim H^{-1}$, while a collider process characterized by an energy scale $E_{\rm coll}$ probes distances and times of order $L_{\rm coll} \sim E_{\rm coll}^{-1}$, then the natural dimensionless expansion parameter controlling
background-gradient corrections is $\epsilon_{\rm bg}
    \equiv
    \frac{L_{\rm coll}}{L_{\rm bg}}
    \sim
\frac{H}{E_{\rm coll}}$. For the present-day cosmological expansion rate, $H_0
\sim 10^{-33}\ {\rm eV}$, and even for a relatively low characteristic collider energy, $E_{\rm coll}
    \sim
    100\ {\rm GeV}
    \sim
10^{11}\ {\rm eV}$, one obtains $\epsilon_{\rm bg} \sim \frac{10^{-33}\ {\rm eV}}{10^{11}\ {\rm eV}} \sim
10^{-44}$. For a TeV-scale process, the corresponding ratio is even smaller,
$\epsilon_{\rm bg}\sim 10^{-45}$. Equivalently, the spatial hierarchy
between the microscopic collider region and the characteristic
cosmological variation scale is $\frac{L_{\rm coll}}{H^{-1}}
    \sim
10^{-44}\text{--}10^{-45}$. The acoustic metric and the coefficients entering the quadratic dark-energy action can therefore be regarded as constant to extremely high accuracy throughout the microscopic spacetime region associated with the collision. More generally, the full acoustic-geometry result may be organized as a derivative expansion around the locally constant background. Denoting
the decay rate computed in the full acoustic geometry by
$\Gamma_{\rm acoustic}$ and its constant-background limit by $\Gamma_{\rm local}$, one expects schematically $\Gamma_{\rm acoustic} =
    \Gamma_{\rm local}
    \left[
        1
        +
        \mathcal{O}\!\left(
            \frac{H}{E_{\rm coll}}
        \right)
        +
        \mathcal{O}\!\left(
            \frac{\dot{c}_s}
                 {E_{\rm coll}c_s}
        \right)
        +
        \mathcal{O}\!\left(
            \frac{\dot{\Sigma}}
                 {E_{\rm coll}\Sigma}
        \right)
        +
        \cdots
\right]$. Here the additional terms represent corrections arising from temporal
or spatial gradients of the background quantities that determine the effective acoustic geometry. If these quantities evolve on a
cosmological time scale, such that parametrically $\frac{\dot{c}_s}{c_s}
\sim H$ and $\frac{\dot{\Sigma}}{\Sigma}
\sim H$, then all such corrections are suppressed by powers of $H/E_{\rm coll}$. Under the homogeneous late-time cosmological background assumed in the collider analysis, the leading result
therefore satisfies parametrically $\Gamma_{\rm acoustic}
    =
    \Gamma_{\rm collider\,EFT}
    \left[
        1
        +
        \mathcal{O}\!\left(
            \frac{H_0}{E_{\rm coll}}
        \right)
\right]$ and hence, for $E_{\rm coll}\gtrsim100~{\rm GeV}$, $\Gamma_{\rm acoustic}
    =
    \Gamma_{\rm collider\,EFT}
    \left[
        1+\mathcal{O}(10^{-44})
\right]$, up to model-dependent microscopic corrections associated with the
validity and truncation of the underlying EFT. Thus, within the separation of scales assumed here, the full acoustic-geometry description reduces locally to the constant-background collider EFT to extremely high accuracy. The two calculations are therefore not merely mutually consistent in this regime: the collider result is the leading term in the local derivative expansion of the acoustic-geometry description.

\section{Discussion and Conclusions}

In this work we have developed a unified effective-field-theory framework in which collider measurements of BSM mediator resonances provide a direct probe of dark energy microphysics. 
By embedding a dynamical dark energy scalar sector into the 2HDM+$a$ model through symmetry-motivated derivative interactions, we have shown that the properties of a pseudoscalar mediator resonance—its decay widths, branching ratios, and kinematic structure—can become sensitive to parameters that are traditionally associated with cosmological dynamics.

A central result of this work is that BSM mediator observables at colliders encode information about three key quantities that characterize dark energy microphysics. 
First, the kinetic normalization factor $Z_a$, induced by higher-dimensional operators, captures the influence of the background dark energy kinetic density on the propagation and normalization of the mediator field. 
This effect leads to correlated shifts in the physical mediator mass, couplings, and kinematic thresholds, thereby modifying resonance lineshapes in a way that is directly testable in collider experiments. 
We stress that the conservative cosmological-background expectation corresponds to $Z_a\simeq1$. The $Z_a\neq1$ regime explored here instead provides a model-independent parameterization of possible enhanced local kinetic backgrounds. A fully predictive realization of this regime would require specifying the mechanism that establishes the local scalar profile and, where relevant, confronting its matter couplings with laboratory, and Solar-System constraints. Second, the parameter $\kappa = P_X$ controls the normalization of dark energy fluctuations and enters the decay widths through inverse powers, implying that even $\mathcal{O}(1)$ variations can produce parametrically distinct collider signatures. 
Third, and most importantly, the sound speed of dark energy $c_s^2$ governs the propagation of fluctuations and introduces a characteristic dependence in the invisible decay width through both the matrix element and the modified phase-space measure. 
The resulting scaling, $(1+c_s^2)^2/c_s^3$, implies that subluminal sound speeds can significantly enhance invisible decay channels and alter the observable properties of the mediator resonance.

These results highlight a particularly important and timely opportunity. 
While cosmological observations have constrained the dark energy equation-of-state parameter $w$ with impressive precision~\cite{Planck2018}, this constraint alone does not uniquely determine the underlying microphysics~\cite{Gurrola:2026fqx,Gurrola:2026nks}. 
A wide range of theoretically distinct models -- including canonical quintessence, k-essence, coupled dark energy, and more exotic scenarios -- can produce nearly identical expansion histories characterized by similar values of $w$. 
In contrast, the sound speed $c_s^2$ provides a far more discriminating observable, as it directly controls the clustering and propagation properties of dark energy perturbations. 
However, $c_s^2$ remains only weakly constrained by current cosmological and astrophysical probes, which are primarily sensitive to integrated effects over large scales and long times.

In this context, collider measurements of BSM mediator resonances offer a qualitatively new and complementary probe, being sensitive to the local behavior of dark energy fluctuations through direct couplings to mediator fields. 
As demonstrated in this work, the dependence of decay widths and kinematics on $c_s^2$ provides a novel pathway to access this otherwise elusive parameter. 
This establishes the possibility that precision measurements of BSM mediator resonances at colliders like the LHC could discriminate between different classes of dark energy models that are otherwise indistinguishable at the level of background cosmology. More generally, we reveal a deep connection between BSM physics at colliders and the microphysical origin of cosmic acceleration, an intersection of physics which is largely uncharted as of now.

The implications of this framework extend beyond collider phenomenology. 
The same interactions that modify mediator properties at colliders can influence a wide range of cosmological and astrophysical phenomena. 
For example, the coupling between dark energy dynamics and additional fields may affect the evolution of early-Universe phase transitions, potentially altering the spectrum of stochastic gravitational-wave backgrounds accessible to current and future experiments. 
Furthermore, the derivative structure introduced here provides a concrete field-theoretic realization of energy exchange between dark matter and dark energy, a possibility that has been extensively explored in phenomenological studies but rarely grounded in a UV-consistent EFT framework. 
This opens the door to a more systematic investigation of interacting dark-sector models from first principles.

We further note that recent indications of time-dependent dark energy, as reported in Ref.~\cite{DESI:2025zgx}, strengthen the motivation for probes complementary to measurements of the background expansion. We emphasize, however, that the benchmark considered here is not intended to provide a complete cosmological fit to a potentially phantom-like reconstruction of $w(z)$. For the collider phenomenology, the relevant cosmological quantities are those characterizing the present-day dark energy background, including $w_0\equiv w(z=0)$, the dark-energy density, and the perturbative sound speed $c_s^2$. By contrast, the full redshift dependence of $w(z)$ inferred from cosmological observations characterizes the evolution of the dark-energy sector over cosmic time. The collider EFT therefore need not reproduce the entire DESI-preferred $w(z)$ trajectory for the proposed collider dark energy connection to remain meaningful. Rather, it probes the local kinetic and perturbative properties associated with a viable present-day dark-energy background. Moreover, effective phantom-like behavior can arise in broader cosmological realizations into which the derivative interactions studied here may be embedded. Constructing a complete realization that reproduces the inferred expansion history while simultaneously satisfying theoretical stability and observational constraints is beyond the scope of the present work and is left for future study.

Finally, the framework considered here should be understood as a low-energy collider EFT rather than as a complete ultraviolet theory of quintessence. The derivative portal interactions preserve the approximate shift symmetry of the dark-energy field and therefore avoid introducing generic hard breaking of the symmetry responsible for protecting its ultralight nature. Nevertheless, a complete realization must explain the origin and radiative stability of the small quintessence potential, control additional symmetry-breaking effects, and specify the heavy dynamics generating the matching scales $\Lambda_5$ and $\Lambda_8$. The collider analysis is consequently restricted to the perturbative regime below these scales, while constructing a fully natural ultraviolet completion is left for future work.

Looking ahead, our results motivate a broad and interdisciplinary research program. 
On the theoretical side, it will be important to explore ultraviolet completions, investigate the interplay with screening mechanisms, and extend the analysis to other classes of dark energy EFTs. 
On the experimental side, we emphasize that precision studies of BSM mediator resonances represent a promising and largely unexplored avenue for probing dark energy microphysics. 
To date, collider searches have not been interpreted within this context, and the possibility that resonance measurements could encode information about cosmic acceleration remains largely untapped.

We therefore strongly encourage the collider physics community to pursue this direction. 
Dedicated analyses targeting deviations in resonance lineshapes, branching ratios, and missing-energy signatures could provide the first experimental constraints on the sound speed and kinetic structure of dark energy. 
The prospect that measurements of BSM mediator resonances at the highest achievable energies could shed light on the physics driving the accelerated expansion of the Universe represents a compelling and unifying goal, bridging cosmology and particle physics in a fundamentally new way.

\noindent \textbf{Acknowledgements:} A.G. and C.B. gratefully acknowledge support from Vanderbilt University and the U.S. National Science Foundation. This work is supported in part by NSF Award PHY-2411502. A.F. and C.R. would like to thank the constant and enduring financial support received for this project from the Physics Department and the Faculty of Science at Universidad de Los Andes (Bogot\'a, Colombia). The authors would like to thank Denis Rathjens for fruitful discussions and valuable suggestions related to this work.

\bibliographystyle{elsarticle-num}
\bibliography{references}

@article{Cheung:2007st,
  author = "Cheung, Clifford and Creminelli, Paolo and Fitzpatrick, A. Liam and Kaplan, Jared and Senatore, Leonardo",
  title = "{The Effective Field Theory of Inflation}",
  journal = "JHEP",
  volume = "03",
  pages = "014",
  year = "2008",
  eprint = "0709.0293",
  archivePrefix = "arXiv",
  primaryClass = "hep-th",
  doi = "10.1088/1126-6708/2008/03/014"
}

@article{perlmutter1999measurements,
  author = {Perlmutter, S. et al.},
  title = {Measurements of Omega and Lambda from 42 High-Redshift Supernovae},
  journal = {Astrophys. J.},
  volume = {517},
  pages = {565--586},
  year = {1999},
  doi = {10.1086/307221}
}

@article{weinberg2013observational,
  author = {Weinberg, D. H. and Mortonson, M. J. and Eisenstein, D. J. and Hirata, C. and Riess, A. G. and Rozo, E.},
  title = {Observational Probes of Cosmic Acceleration},
  journal = {Phys. Rept.},
  volume = {530},
  pages = {87--255},
  year = {2013},
  doi = {10.1016/j.physrep.2013.05.001}
}

@article{copeland2006dynamics,
  author = {Copeland, E. J. and Sami, M. and Tsujikawa, S.},
  title = {Dynamics of Dark Energy},
  journal = {Int. J. Mod. Phys. D},
  volume = {15},
  pages = {1753--1936},
  year = {2006},
  doi = {10.1142/S021827180600942X}
}

@book{amendola2010dark,
  author = {Amendola, L. and Tsujikawa, S.},
  title = {Dark Energy: Theory and Observations},
  publisher = {Cambridge University Press},
  year = {2010},
  doi = {10.1017/CBO9780511750823}
}

@article{atlas2012observation,
  author = {{ATLAS Collaboration}},
  title = {Observation of a New Particle in the Search for the Standard Model Higgs Boson with the ATLAS Detector at the LHC},
  journal = {Phys. Lett. B},
  volume = {716},
  pages = {1--29},
  year = {2012},
  doi = {10.1016/j.physletb.2012.08.020}
}

@article{cms2012observation,
  author = {{CMS Collaboration}},
  title = {Observation of a New Boson at a Mass of 125 GeV with the CMS Experiment at the LHC},
  journal = {Phys. Lett. B},
  volume = {716},
  pages = {30--61},
  year = {2012},
  doi = {10.1016/j.physletb.2012.08.021}
}

@article{carroll1998quintessence,
  author = {Carroll, S. M.},
  title = {Quintessence and the Rest of the World: Suppressing Long-Range Interactions},
  journal = {Phys. Rev. Lett.},
  volume = {81},
  pages = {3067--3070},
  year = {1998},
  doi = {10.1103/PhysRevLett.81.3067}
}

@article{ferreira1998quintessence,
  author = {Ferreira, P. G. and Joyce, M.},
  title = {Cosmology with a Primordial Scaling Field},
  journal = {Phys. Rev. D},
  volume = {58},
  pages = {023503},
  year = {1998},
  doi = {10.1103/PhysRevD.58.023503}
}

@article{brax2004detecting,
  author = {Brax, P. and van de Bruck, C. and Davis, A.-C. and Khoury, J. and Weltman, A.},
  title = {Detecting Dark Energy in Orbit: The Cosmological Chameleon},
  journal = {Phys. Rev. D},
  volume = {70},
  pages = {123518},
  year = {2004},
  doi = {10.1103/PhysRevD.70.123518}
}

@article{burrage2018tests,
  author = {Burrage, C. and Sakstein, J.},
  title = {Tests of Chameleon Gravity},
  journal = {Living Rev. Relativ.},
  volume = {21},
  pages = {1},
  year = {2018},
  doi = {10.1007/s41114-018-0011-x}
}

@article{de1SupernovaSearchTeam:1998fmf,
    author = "Riess, Adam G. and others",
    collaboration = "Supernova Search Team",
    title = "{Observational evidence from supernovae for an accelerating universe and a cosmological constant}",
    eprint = "astro-ph/9805201",
    archivePrefix = "arXiv",
    doi = "10.1086/300499",
    journal = "Astron. J.",
    volume = "116",
    pages = "1009--1038",
    year = "1998"
}

@article{de2Li:2012dt,
    author = "Li, Miao and Li, Xiao-Dong and Wang, Shuang and Wang, Yi",
    title = "{Dark Energy: A Brief Review}",
    eprint = "1209.0922",
    archivePrefix = "arXiv",
    primaryClass = "astro-ph.CO",
    doi = "10.1007/s11467-013-0300-5",
    journal = "Front. Phys. (Beijing)",
    volume = "8",
    pages = "828--846",
    year = "2013"
}

@article{de3Li:2011sd,
    author = "Li, Miao and Li, Xiao-Dong and Wang, Shuang and Wang, Yi",
    title = "{Dark Energy}",
    eprint = "1103.5870",
    archivePrefix = "arXiv",
    primaryClass = "astro-ph.CO",
    reportNumber = "CAS-KITPC-ITP-249",
    doi = "10.1088/0253-6102/56/3/24",
    journal = "Commun. Theor. Phys.",
    volume = "56",
    pages = "525--604",
    year = "2011"
}

@online{Mortonson:2013zfa,
    author = "Mortonson, Michael J. and Weinberg, David H. and White, Martin",
    title = "{Dark Energy: A Short Review}",
    eprint = "1401.0046",
    archivePrefix = "arXiv",
    primaryClass = "astro-ph.CO",
    month = "12",
    year = "2013"
}

@article{de5Frusciante:2019xia,
    author = "Frusciante, Noemi and Perenon, Louis",
    title = "{Effective field theory of dark energy: A review}",
    eprint = "1907.03150",
    archivePrefix = "arXiv",
    primaryClass = "astro-ph.CO",
    doi = "10.1016/j.physrep.2020.02.004",
    journal = "Phys. Rept.",
    volume = "857",
    pages = "1--63",
    year = "2020"
}

@article{de6Huterer:2017buf,
    author = "Huterer, Dragan and Shafer, Daniel L",
    title = "{Dark energy two decades after: Observables, probes, consistency tests}",
    eprint = "1709.01091",
    archivePrefix = "arXiv",
    primaryClass = "astro-ph.CO",
    doi = "10.1088/1361-6633/aa997e",
    journal = "Rept. Prog. Phys.",
    volume = "81",
    number = "1",
    pages = "016901",
    year = "2018"
}

@article{de7Vagnozzi:2021quy,
    author = "Vagnozzi, Sunny and Visinelli, Luca and Brax, Philippe and Davis, Anne-Christine and Sakstein, Jeremy",
    title = "{Direct detection of dark energy: The XENON1T excess and future prospects}",
    eprint = "2103.15834",
    archivePrefix = "arXiv",
    primaryClass = "hep-ph",
    doi = "10.1103/PhysRevD.104.063023",
    journal = "Phys. Rev. D",
    volume = "104",
    number = "6",
    pages = "063023",
    year = "2021"
}

@article{de8Adil:2023ara,
    author = "Adil, Shahnawaz A. and Mukhopadhyay, Upala and Sen, Anjan A. and Vagnozzi, Sunny",
    title = "{Dark energy in light of the early JWST observations: case for a negative cosmological constant?}",
    eprint = "2307.12763",
    archivePrefix = "arXiv",
    primaryClass = "astro-ph.CO",
    doi = "10.1088/1475-7516/2023/10/072",
    journal = "JCAP",
    volume = "10",
    pages = "072",
    year = "2023"
}

@article{de10DiValentino:2020evt,
    author = "Di Valentino, Eleonora and Gariazzo, Stefano and Mena, Olga and Vagnozzi, Sunny",
    title = "{Soundness of Dark Energy properties}",
    eprint = "2005.02062",
    archivePrefix = "arXiv",
    primaryClass = "astro-ph.CO",
    doi = "10.1088/1475-7516/2020/07/045",
    journal = "JCAP",
    volume = "07",
    number = "07",
    pages = "045",
    year = "2020"
}

@article{de11Nojiri:2010wj,
    author = "Nojiri, Shin'ichi and Odintsov, Sergei D.",
    title = "{Unified cosmic history in modified gravity: from F(R) theory to Lorentz non-invariant models}",
    eprint = "1011.0544",
    archivePrefix = "arXiv",
    primaryClass = "gr-qc",
    doi = "10.1016/j.physrep.2011.04.001",
    journal = "Phys. Rept.",
    volume = "505",
    pages = "59--144",
    year = "2011"
}

@article{de12Nojiri:2006ri,
    author = "Nojiri, Shin'ichi and Odintsov, Sergei D.",
    editor = "Borowiec, Andrzej",
    title = "{Introduction to modified gravity and gravitational alternative for dark energy}",
    eprint = "hep-th/0601213",
    archivePrefix = "arXiv",
    reportNumber = "KARP-2006-06",
    doi = "10.1142/S0219887807001928",
    journal = "eConf",
    volume = "C0602061",
    pages = "06",
    year = "2006"
}

@article{csc3ballesteros2010dark,
  title={Dark energy with non-adiabatic sound speed: initial conditions and detectability},
  author={Ballesteros, Guillermo and Lesgourgues, Julien},
  journal={Journal of Cosmology and Astroparticle Physics},
  volume={2010},
  number={10},
  pages={014},
  year={2010},
  publisher={IOP Publishing}
}

@article{csc1Hannestad:2005ak,
    author = "Hannestad, Steen",
    title = "{Constraints on the sound speed of dark energy}",
    eprint = "astro-ph/0504017",
    archivePrefix = "arXiv",
    doi = "10.1103/PhysRevD.71.103519",
    journal = "Phys. Rev. D",
    volume = "71",
    pages = "103519",
    year = "2005"
}

@article{csc2de2010measuring,
  title={Measuring the speed of dark: Detecting dark energy perturbations},
  author={de Putter, Roland and Huterer, Dragan and Linder, Eric V},
  journal={Physical Review D—Particles, Fields, Gravitation, and Cosmology},
  volume={81},
  number={10},
  pages={103513},
  year={2010},
  publisher={APS}
}

@article{csc4linton2018variable,
  title={Variable sound speed in interacting dark energy models},
  author={Linton, Mark S and Pourtsidou, Alkistis and Crittenden, Robert and Maartens, Roy},
  journal={Journal of Cosmology and Astroparticle Physics},
  volume={2018},
  number={04},
  pages={043},
  year={2018},
  publisher={IOP Publishing}
}

@article{csc5bean2004probing,
  title={Probing dark energy perturbations: the dark energy equation of state and speed of sound as measured by WMAP},
  author={Bean, Rachel and Dore, Olivier},
  journal={Physical Review D},
  volume={69},
  number={8},
  pages={083503},
  year={2004},
  publisher={APS}
}

@article{csc6eisenstein2005dark,
  title={Dark energy and cosmic sound},
  author={Eisenstein, Daniel J},
  journal={New Astronomy Reviews},
  volume={49},
  number={7-9},
  pages={360--365},
  year={2005},
  publisher={Elsevier}
}

@article{csc7sergijenko2015sound,
  title={Sound speed of scalar field dark energy: weak effects and large uncertainties},
  author={Sergijenko, Olga and Novosyadlyj, Bohdan},
  journal={Physical Review D},
  volume={91},
  number={8},
  pages={083007},
  year={2015},
  publisher={APS}
}

@online{Shen:2026qxg,
    author = "Shen, Yin-Fa and Gurrola, Alfredo and Romeo, Francesco and Rathjens, Denis and Fl{\'o}rez, Andres",
    title = "{Heavy Neutrinos across the Electroweak-to-Multi-TeV Frontier via Novel ML-Enhanced Probes}",
    eprint = "2601.09095",
    archivePrefix = "arXiv",
    primaryClass = "hep-ph",
    month = "1",
    year = "2026"
}

@online{Shen:2025nkr,
    author = "Shen, Yin-Fa and Gurrola, Alfredo",
    title = "{Large-Width New Physics at Colliders: A Gauge-Invariant Resummation Approach}",
    eprint = "2511.15114",
    archivePrefix = "arXiv",
    primaryClass = "hep-ph",
    month = "11",
    year = "2025"
}

@article{Qureshi:2024cmg,   
    author = "Qureshi, Umar Sohail and Gurrola, Alfredo and Fl{\'o}rez, Andres",
    title = "{Probing compressed mass spectrum supersymmetry at the high-luminosity LHC with the vector boson fusion topology}",
    eprint = "2411.13837",
    archivePrefix = "arXiv",
    primaryClass = "hep-ph",
    doi = "10.1140/epjc/s10052-025-14935-y",
    journal = "Eur. Phys. J. C",
    volume = "85",
    number = "10",
    pages = "1208",
    year = "2025"
}

@article{Qureshi:2024naw,
    author = "Qureshi, Umar Sohail and Gurrola, Alfredo and Fl{\'o}rez, Andres and Rodriguez, Cristian",
    title = "{Probing light scalars and vector-like quarks at the high-luminosity LHC}",
    eprint = "2410.17854",
    archivePrefix = "arXiv",
    primaryClass = "hep-ph",
    doi = "10.1140/epjc/s10052-025-14085-1",
    journal = "Eur. Phys. J. C",
    volume = "85",
    number = "4",
    pages = "379",
    year = "2025"
}

@article{Barbosa:2022mmw,
    author = "Barbosa, Diego and D{\'\i}az, Felipe and Quintero, Liliana and Fl{\'o}rez, Andr{\'e}s and Sanchez, Manuel and Gurrola, Alfredo and Sheridan, Elijah and Romeo, Francesco",
    title = "{Probing a $\textrm{Z}^{\prime }$ with non-universal fermion couplings through top quark fusion, decays to bottom quarks, and machine learning techniques}",
    eprint = "2210.15813",  
    archivePrefix = "arXiv",
    primaryClass = "hep-ph",
    doi = "10.1140/epjc/s10052-023-11506-x",
    journal = "Eur. Phys. J. C",
    volume = "83", 
    number = "5",
    pages = "413",
    year = "2023"
}

@inproceedings{Gurrola:2022ssc,
    author = "Gurrola, Alfredo and Ruiz-{\'A}lvarez, Jos{\'e} David",
    title = "{Quarkophobic W' for LHC searches}",
    booktitle = "{20th Conference on Flavor Physics and CP Violation~}",
    eprint = "2208.01861",
    archivePrefix = "arXiv",
    primaryClass = "hep-ph",
    month = "8", 
    year = "2022" 
}

@article{Dutta:2022bfq,
    author = "Dutta, Bhaskar and Ghosh, Sumit and Gurrola, Alfredo and Julson, Dale and Kamon, Teruki and Kumar, Jason",
    title = "{Probing an MeV-scale scalar boson in association with a TeV-Scale top-quark partner at the LHC}",
    eprint = "2202.08234",
    archivePrefix = "arXiv",
    primaryClass = "hep-ph",
    reportNumber = "APS/123-QED, MI-TH-771, KIAS-P22060",
    doi = "10.1007/JHEP03(2023)164",
    journal = "JHEP",
    volume = "03", 
    pages = "164",
    year = "2023" 
}

@article{Cardona:2021ebw,
    author = "Cardona, Nathalia and Andr{\'e}s, Fl{\'o}rez and Alfredo, Gurrola and Will, Johns and Paul, Sheldon and cheng, Tao",
    title = "{Long-term LHC discovery reach for compressed Supersymmetry models using VBF processes}",
    eprint = "2102.10194",
    archivePrefix = "arXiv",
    primaryClass = "hep-ph",
    doi = "10.1007/JHEP11(2022)026",
    journal = "JHEP",
    volume = "11",
    pages = "026",
    year = "2022"
}

@article{Florez:2021zoo,
    author = "Fl{\'o}rez, Andr{\'e}s and Gurrola, Alfredo and Johns, Will and Sheldon, Paul and Sheridan, Elijah and Sinha, Kuver and Soubasis, Brandon",
    title = "{Probing axionlike particles with $\gamma\gamma$ final states from vector boson fusion processes at the LHC}",
    eprint = "2101.11119",  
    archivePrefix = "arXiv",
    primaryClass = "hep-ph",
    doi = "10.1103/PhysRevD.103.095001",
    journal = "Phys. Rev. D",
    volume = "103", 
    number = "9", 
    pages = "095001",
    year = "2021"
}

@article{Florez:2019tqr,
    author = "Fl{\'o}rez, Andr{\'e}s and Gurrola, Alfredo and Johns, Will and Maruri, Jessica and Sheldon, Paul and Sinha, Kuver and Starko, Savanna Rae",
    title = "{Anapole Dark Matter via Vector Boson Fusion Processes at the LHC}",
    eprint = "1902.01488",
    archivePrefix = "arXiv",
    primaryClass = "hep-ph",
    doi = "10.1103/PhysRevD.100.016017",
    journal = "Phys. Rev. D",
    volume = "100",
    number = "1",
    pages = "016017",
    year = "2019"
}

@article{Florez:2018ojp,
    author = "Fl{\'o}rez, Andr{\'e}s and Guo, Yuhan and Gurrola, Alfredo and Johns, Will and Ray, Oishik and Sheldon, Paul and Starko, Savanna",
    title = "{Probing heavy spin-2 bosons with $\gamma\gamma$ final states from vector boson fusion processes at the LHC}",
    eprint = "1812.06824",
    archivePrefix = "arXiv",
    primaryClass = "hep-ph",
    doi = "10.1103/PhysRevD.99.035034",
    journal = "Phys. Rev. D",
    volume = "99",
    number = "3",
    pages = "035034",
    year = "2019"
}

@article{Leonardi:2018jzn,
    author = "Leonardi, Roberto and Panella, Orlando and Romeo, Francesco and Gurrola, Alfredo and Sun, Hao and Xue, She-Sheng",
    title = "{Phenomenology at the LHC of composite particles from strongly interacting Standard Model fermions via four-fermion operators of NJL type}",
    eprint = "1810.11420",
    archivePrefix = "arXiv",
    primaryClass = "hep-ph",
    doi = "10.1140/epjc/s10052-020-7822-0",
    journal = "Eur. Phys. J. C",
    volume = "80",
    number = "4",
    pages = "309",
    year = "2020"
}

@article{Avila:2018sja,
    author = "Avila, Carlos and Fl{\'o}rez, Andr{\'e}s and Gurrola, Alfredo and Julson, Dale and Starko, Savanna",
    title = "{Connecting particle physics and cosmology: Measuring the dark matter relic density in compressed supersymmetry models at the LHC}",
    eprint = "1801.03966",  
    archivePrefix = "arXiv",
    primaryClass = "hep-ph", 
    doi = "10.1016/j.dark.2019.100430",
    journal = "Phys. Dark Univ.",
    volume = "27",   
    pages = "100430",
    year = "2020"
}

@article{Florez:2017xhf,
    author = "Fl{\'o}rez, Andr{\'e}s and Gui, Kaiwen and Gurrola, Alfredo and Pati{\~n}o, Carlos and Restrepo, Diego",
    title = "{Expanding the Reach of Heavy Neutrino Searches at the LHC}",
    eprint = "1708.03007",  
    archivePrefix = "arXiv",
    primaryClass = "hep-ph",
    doi = "10.1016/j.physletb.2018.01.009",
    journal = "Phys. Lett. B",
    volume = "778",
    pages = "94--100",
    year = "2018"
}

@article{Florez:2016uob,
    author = "Fl{\'o}rez, Andr{\'e}s and Gurrola, Alfredo and Johns, Will and Oh, Young Do and Sheldon, Paul and Teague, Dylan and Weiler, Thomas",
    title = "{Searching for New Heavy Neutral Gauge Bosons using Vector Boson Fusion Processes at the LHC}",
    eprint = "1609.09765",  
    archivePrefix = "arXiv",
    primaryClass = "hep-ph", 
    doi = "10.1016/j.physletb.2017.01.062",
    journal = "Phys. Lett. B",   
    volume = "767",  
    pages = "126--132",
    year = "2017"
}

@article{Florez:2016lwi,
    author = "Fl{\'o}rez, Andr{\'e}s and Bravo, Luis and Gurrola, Alfredo and {\'A}vila, Carlos and Segura, Manuel and Sheldon, Paul and Johns, Will",
    title = "{Probing the stau-neutralino coannihilation region at the LHC with a soft tau lepton and a jet from initial state radiation}",
    eprint = "1606.08878",  
    archivePrefix = "arXiv",
    primaryClass = "hep-ph",
    doi = "10.1103/PhysRevD.94.073007",
    journal = "Phys. Rev. D", 
    volume = "94", 
    number = "7",
    pages = "073007",
    year = "2016"
}

@article{Dutta:2015hra,
    author = "Dutta, Bhaskar and Gurrola, Alfredo and Hatakeyama, Kenichi and Johns, Will and Kamon, Teruki and Sheldon, Paul and Sinha, Kuver and Wu, Sean and Wu, Zhenbin",
    title = "{Probing Compressed Bottom Squarks with Boosted Jets and Shape Analysis}",
    eprint = "1507.01001",  
    archivePrefix = "arXiv", 
    primaryClass = "hep-ph",
    doi = "10.1103/PhysRevD.92.095009",
    journal = "Phys. Rev. D",
    volume = "92",
    number = "9",
    pages = "095009",
    year = "2015"
}

@article{Dutta:2014jda,
    author = "Dutta, Bhaskar and Ghosh, Tathagata and Gurrola, Alfredo and Johns, Will and Kamon, Teruki and Sheldon, Paul and Sinha, Kuver and Wang, Kechen and Wu, Sean",
    title = "{Probing Compressed Sleptons at the LHC using Vector Boson Fusion Processes}",
    eprint = "1411.6043",   
    archivePrefix = "arXiv",
    primaryClass = "hep-ph",  
    reportNumber = "MIFPA-14-33",
    doi = "10.1103/PhysRevD.91.055025",
    journal = "Phys. Rev. D",
    volume = "91",
    number = "5",
    pages = "055025",
    year = "2015"
}

@article{Dutta:2013gga,     
    author = "Dutta, Bhaskar and Flanagan, Will and Gurrola, Alfredo and Johns, Will and Kamon, Teruki and Sheldon, Paul and Sinha, Kuver and Wang, Kechen and Wu, Sean",
    title = "{Probing compressed top squark scenarios at the LHC at 14 TeV}",
    eprint = "1312.1348",
    archivePrefix = "arXiv", 
    primaryClass = "hep-ph",
    doi = "10.1103/PhysRevD.90.095022",
    journal = "Phys. Rev. D",
    volume = "90",
    number = "9",
    pages = "095022",
    year = "2014"
}

@article{Dutta:2012xe,      
    author = "Dutta, Bhaskar and Gurrola, Alfredo and Johns, Will and Kamon, Teruki and Sheldon, Paul and Sinha, Kuver",
    title = "{Vector Boson Fusion Processes as a Probe of Supersymmetric Electroweak Sectors at the LHC}",
    eprint = "1210.0964",
    archivePrefix = "arXiv",
    primaryClass = "hep-ph", 
    doi = "10.1103/PhysRevD.87.035029",
    journal = "Phys. Rev. D",
    volume = "87",   
    number = "3",
    pages = "035029",
    year = "2013"
}

@article{Dutta:2008ge,
    author = "Dutta, Bhaskar and Gurrola, Alfredo and Kamon, Teruki and Krislock, Abram and Lahanas, A. B. and Mavromatos, N. E. and Nanopoulos, D. V.",
    title = "{Supersymmetry Signals of Supercritical String Cosmology at the Large Hadron Collider}",
    eprint = "0808.1372",
    archivePrefix = "arXiv",
    primaryClass = "hep-ph",
    reportNumber = "MIFP-08-20, ACT-04-08",
    doi = "10.1103/PhysRevD.79.055002",
    journal = "Phys. Rev. D",
    volume = "79",
    pages = "055002",
    year = "2009"
}

@article{Arnowitt:2008bz,
    author = "Arnowitt, Richard L. and Dutta, Bhaskar and Gurrola, Alfredo and Kamon, Teruki and Krislock, Abram and Toback, David",
    title = "{Determining the Dark Matter Relic Density in the mSUGRA ({\textasciitilde}X0(1))-{\textasciitilde}tau Co-Annhiliation Region at the LHC}",
    eprint = "0802.2968",  
    archivePrefix = "arXiv",
    primaryClass = "hep-ph",
    reportNumber = "MIFP-08-02",
    doi = "10.1103/PhysRevLett.100.231802",
    journal = "Phys. Rev. Lett.",
    volume = "100",
    pages = "231802",
    year = "2008"
}

@article{Delannoy:2013ata,
    author = "Delannoy, Andres G. and others",
    title = "{Probing Dark Matter at the LHC using Vector Boson Fusion Processes}",
    eprint = "1304.7779",
    archivePrefix = "arXiv",
    primaryClass = "hep-ph",
    doi = "10.1103/PhysRevLett.111.061801",
    journal = "Phys. Rev. Lett.",
    volume = "111",
    pages = "061801",
    year = "2013"
}

@article{Branco:2011iw,
    author = "Branco, G. C. and Ferreira, P. M. and Lavoura, L. and Rebelo, M. N. and Sher, Marc and Silva, Joao P.",
    title = "{Theory and phenomenology of two-Higgs-doublet models}",
    eprint = "1106.0034",
    archivePrefix = "arXiv",
    primaryClass = "hep-ph",
    doi = "10.1016/j.physrep.2012.02.002",
    journal = "Phys. Rept.",
    volume = "516",
    pages = "1--102",
    year = "2012"
}

@online{Craig:2013hca,
    author = "Craig, Nathaniel and Galloway, Jamison and Thomas, Scott",
    title = "{Searching for Signs of the Second Higgs Doublet}",
    eprint = "1305.2424",
    archivePrefix = "arXiv",
    primaryClass = "hep-ph",
    reportNumber = "RU-NHETC-2013-07",
    month = "5",
    year = "2013"
}

@article{Gunion:2002zf,
    author = "Gunion, John F. and Haber, Howard E.",
    title = "{The CP conserving two Higgs doublet model: The Approach to the decoupling limit}",
    eprint = "hep-ph/0207010",
    archivePrefix = "arXiv",
    reportNumber = "SCIPP-02-10",
    doi = "10.1103/PhysRevD.67.075019",
    journal = "Phys. Rev. D",
    volume = "67",
    pages = "075019",
    year = "2003"
}

@article{Abercrombie:2015wmb,
    author = "Abercrombie, Daniel and others",
    editor = "Boveia, Antonio and Doglioni, Caterina and Lowette, Steven and Malik, Sarah and Mrenna, Stephen",
    title = "{Dark Matter benchmark models for early LHC Run-2 Searches: Report of the ATLAS/CMS Dark Matter Forum}",
    eprint = "1507.00966",
    archivePrefix = "arXiv",
    primaryClass = "hep-ex",
    reportNumber = "FERMILAB-PUB-15-282-CD",
    doi = "10.1016/j.dark.2019.100371",
    journal = "Phys. Dark Univ.",
    volume = "27",
    pages = "100371",
    year = "2020"
}

@article{Abdallah:2015ter,
    author = "Abdallah, Jalal and others",
    title = "{Simplified Models for Dark Matter Searches at the LHC}",
    eprint = "1506.03116",
    archivePrefix = "arXiv",
    primaryClass = "hep-ph",
    reportNumber = "FERMILAB-PUB-15-283-CD, CERN-PH-TH-2015-139",
    doi = "10.1016/j.dark.2015.08.001",
    journal = "Phys. Dark Univ.",
    volume = "9-10",
    pages = "8--23",
    year = "2015"
}

@article{Boveia:2018yeb,
    author = "Boveia, Antonio and Doglioni, Caterina",
    title = "{Dark Matter Searches at Colliders}",
    eprint = "1810.12238",
    archivePrefix = "arXiv",
    primaryClass = "hep-ex",
    doi = "10.1146/annurev-nucl-101917-021008",
    journal = "Ann. Rev. Nucl. Part. Sci.",
    volume = "68",
    pages = "429--459",
    year = "2018"
}

@article{CMS:2025kzt,
    author = "Hayrapetyan, Aram and others",
    collaboration = "CMS",
    title = "{Observation of a pseudoscalar excess at the top quark pair production threshold}",
    eprint = "2503.22382",
    archivePrefix = "arXiv",
    primaryClass = "hep-ex",
    reportNumber = "CMS-TOP-24-007, CERN-EP-2025-061",
    doi = "10.1088/1361-6633/adf7d3",
    journal = "Rept. Prog. Phys.",
    volume = "88",
    number = "8",
    pages = "087801",
    year = "2025"
}

@online{ATLAS:2026dbe,
    author = "Aad, Georges and others",
    collaboration = "ATLAS",
    title = "{Observation of a cross-section enhancement near the $t\bar{t}$ production threshold in $\sqrt{s}=13$ TeV $pp$ collisions with the ATLAS detector}",
    eprint = "2601.11780",
    archivePrefix = "arXiv",
    primaryClass = "hep-ex",
    reportNumber = "CERN-EP-2026-002",
    month = "1",
    year = "2026"
}

@article{Bauer:2017ota,
  author = {Bauer, Martin and Haisch, Ulrich and Kahlhoefer, Felix},
  title = {Simplified dark matter models with two Higgs doublets: I. Pseudoscalar mediators},
  journal = {JHEP},
  volume = {05},
  pages = {138},
  year = {2017},
  eprint = {1701.07427},
  archivePrefix = {arXiv},
  primaryClass = {hep-ph}
}

@article{Bauer:2017nlg,
  author = {Bauer, Martin and Haisch, Ulrich and Kahlhoefer, Felix},
  title = {Simplified dark matter models with two Higgs doublets: II. Phenomenology and constraints},
  journal = {JHEP},
  volume = {05},
  pages = {103},
  year = {2017},
  eprint = {1701.07427},
  archivePrefix = {arXiv},
  primaryClass = {hep-ph}
}

@article{No:2015xqa,
  author = {No, Jose M.},
  title = {Looking for Higgs portal dark matter with mono-Higgs signatures at the LHC},
  journal = {Phys. Rev. D},
  volume = {93},
  pages = {031701},
  year = {2016},
  eprint = {1509.01110},
  archivePrefix = {arXiv},
  primaryClass = {hep-ph}
}

@article{Arina:2016cqj,
  author = {Arina, Chiara and Backovic, Mihailo and Heisig, Jan and Maltoni, Fabio and Pellen, Mathieu},
  title = {A comprehensive analysis of dark matter models with pseudoscalar mediators},
  journal = {JHEP},
  volume = {11},
  pages = {111},
  year = {2016},
  eprint = {1605.09242},
  archivePrefix = {arXiv},
  primaryClass = {hep-ph}
}

@article{ArmendarizPicon:1999rj,
  author = {Armendariz-Picon, C. and Damour, T. and Mukhanov, V.},
  title = {k-Inflation},
  journal = {Phys. Lett. B},
  volume = {458},
  pages = {209-218},
  year = {1999},
  eprint = {hep-th/9904075}
}

@article{Garriga:1999vw,
  author = {Garriga, Jaume and Mukhanov, Viatcheslav F.},
  title = {Perturbations in k-inflation},
  journal = {Phys. Lett. B},
  volume = {458},
  pages = {219-225},
  year = {1999},
  eprint = {hep-th/9904176}
}

@review{Tsujikawa:2013fta,
  author = {Tsujikawa, Shinji},
  title = {Quintessence: A Review},
  journal = {Class. Quant. Grav.},
  volume = {30},
  pages = {214003},
  year = {2013},
  eprint = {1304.1961},
  archivePrefix = {arXiv},
  primaryClass = {gr-qc}
}

@article{Ratra:1987rm,
  author = {Ratra, Bharat and Peebles, P. J. E.},
  title = {Cosmological Consequences of a Rolling Homogeneous Scalar Field},
  journal = {Phys. Rev. D},
  volume = {37},
  pages = {3406},
  year = {1988}
}

@article{Caldwell:1997ii,
  author = {Caldwell, Robert R. and Dave, Rahul and Steinhardt, Paul J.},
  title = {Cosmological imprint of an energy component with general equation of state},
  journal = {Phys. Rev. Lett.},
  volume = {80},
  pages = {1582-1585},
  year = {1998},
  eprint = {astro-ph/9708069}
}

@article{ArmendarizPicon:2000dh,
  author = {Armendariz-Picon, C. and Mukhanov, V. and Steinhardt, Paul J.},
  title = {A Dynamical solution to the problem of a small cosmological constant and late time cosmic acceleration},
  journal = {Phys. Rev. Lett.},
  volume = {85},
  pages = {4438-4441},
  year = {2000},
  eprint = {astro-ph/0004134}
}

@article{DeDeo:2003te,
  author = {DeDeo, Simon and Caldwell, Robert R. and Steinhardt, Paul J.},
  title = {Effects of the sound speed of quintessence on the microwave background and large scale structure},
  journal = {Phys. Rev. D},
  volume = {67},
  pages = {103509},
  year = {2003},
  eprint = {astro-ph/0301284}
}

@article{Erickson:2001bq,
  author = {Erickson, J. K. and Caldwell, Robert R. and Steinhardt, Paul J. and Armendariz-Picon, C. and Mukhanov, V.},
  title = {Measuring the speed of sound of quintessence},
  journal = {Phys. Rev. Lett.},
  volume = {88},
  pages = {121301},
  year = {2002},
  eprint = {astro-ph/0112438}
}

@article{Arcadi:2017kky,
  author = {Arcadi, Giorgio and Dutra, Mauro and Ghosh, Poulomi and Lindner, Manfred and Mambrini, Yann and Pierre, Mathias and Profumo, Stefano and Queiroz, Farinaldo S.},
  title = {The Waning of the WIMP? A Review of Models, Searches, and Constraints},
  journal = {Eur. Phys. J. C},
  volume = {78},
  pages = {203},
  year = {2018},
  eprint = {1703.07364},
  archivePrefix = {arXiv},
  primaryClass = {hep-ph}
}

@article{arcadi2hdmapheno,
    author = {Arcadi, Giorgio and Benincasa, Nico and Djouadi, Abdelhak and Kannike, Kristjan},
    title = {Two-Higgs-doublet-plus-pseudoscalar model: Collider, dark matter, and gravitational wave signals},
    journal = {Physical Review D},
    volume = {108},
    pages = {055010},
    year = {2023},
    publisher = {APS}
}

@article{cde1gumjudpai2005coupled,
  title={Coupled dark energy: towards a general description of the dynamics},
  author={Gumjudpai, Burin and Naskar, Tapan and Sami, M and Tsujikawa, Shinji},
  journal={Journal of Cosmology and Astroparticle Physics},
  volume={2005},
  number={06},
  pages={007},
  year={2005},
  publisher={IOP Publishing}
}

@article{cde2gomez2020update,
  title={Update on coupled dark energy and the H 0 tension},
  author={G{\'o}mez-Valent, Adri{\`a} and Pettorino, Valeria and Amendola, Luca},
  journal={Physical Review D},
  volume={101},
  number={12},
  pages={123513},
  year={2020},
  publisher={APS}
}

@article{cde3barros2019kinetically,
  title={Kinetically coupled dark energy},
  author={Barros, Bruno J},
  journal={Physical Review D},
  volume={99},
  number={6},
  pages={064051},
  year={2019},
  publisher={APS}
}

@article{cde4xia2009constraint,
  title={Constraint on coupled dark energy models from observations},
  author={Xia, Jun-Qing},
  journal={Physical Review D—Particles, Fields, Gravitation, and Cosmology},
  volume={80},
  number={10},
  pages={103514},
  year={2009},
  publisher={APS}
}

@article{cde5maccio2004coupled,
  title={Coupled dark energy: Parameter constraints from N-body simulations},
  author={Macci{\`o}, Andrea V and Quercellini, Claudia and Mainini, Roberto and Amendola, Luca and Bonometto, Silvio A},
  journal={Physical Review D},
  volume={69},
  number={12},
  pages={123516},
  year={2004},
  publisher={APS}
}

@article{eft1gubitosi2013effective,
  title={The effective field theory of dark energy},
  author={Gubitosi, Giulia and Piazza, Federico and Vernizzi, Filippo},
  journal={Journal of Cosmology and Astroparticle Physics},
  volume={2013},
  number={02},
  pages={032},
  year={2013},
  publisher={IOP Publishing}
}

@article{eft2linder2016effective,
  title={Is the effective field theory of dark energy effective?},
  author={Linder, Eric V and Seng{\"o}r, Gizem and Watson, Scott},
  journal={Journal of Cosmology and Astroparticle Physics},
  volume={2016},
  number={05},
  pages={053},
  year={2016},
  publisher={IOP Publishing}
}

@article{eft3bloomfield2013dark,
  title={Dark energy or modified gravity? An effective field theory approach},
  author={Bloomfield, Jolyon and Flanagan, {\'E}anna {\'E} and Park, Minjoon and Watson, Scott},
  journal={Journal of Cosmology and Astroparticle Physics},
  volume={2013},
  number={08},
  pages={010},
  year={2013},
  publisher={IOP Publishing}
}

@article{eft4liang2023dark,
  title={Dark sector effective field theory},
  author={Liang, Jin-Han and Liao, Yi and Ma, Xiao-Dong and Wang, Hao-Lin},
  journal={Journal of High Energy Physics},
  volume={2023},
  number={12},
  pages={1--36},
  year={2023},
  publisher={Springer}
}

@article{eft5frusciante2014effective,
  title={Effective field theory of dark energy: a dynamical analysis},
  author={Frusciante, Noemi and Raveri, Marco and Silvestri, Alessandra},
  journal={Journal of Cosmology and Astroparticle Physics},
  volume={2014},
  number={02},
  pages={026},
  year={2014},
  publisher={IOP Publishing}
}

@article{nc1armendariz2001essentials,
  title={Essentials of k-essence},
  author={Armendariz-Picon, Christian and Mukhanov, Viatcheslav and Steinhardt, Paul J},
  journal={Physical Review D},
  volume={63},
  number={10},
  pages={103510},
  year={2001},
  publisher={APS}
}

@article{nc2malquarti2003new,
  title={A new view of k-essence},
  author={Malquarti, Michael and Copeland, Edmund J and Liddle, Andrew R and Trodden, Mark},
  journal={Physical Review D},
  volume={67},
  number={12},
  pages={123503},
  year={2003},
  publisher={APS}
}

@article{nc3armendariz2005haloes,
  title={Haloes of k-essence},
  author={Armendariz-Picon, Christian and Lim, Eugene A},
  journal={Journal of Cosmology and Astroparticle Physics},
  volume={2005},
  number={08},
  pages={007},
  year={2005},
  publisher={IOP Publishing}
}

@article{nc4ahn2009dark,
  title={Dark energy properties in DBI theory},
  author={Ahn, Changrim and Kim, Chanju and Linder, Eric V},
  journal={Physical Review D—Particles, Fields, Gravitation, and Cosmology},
  volume={80},
  number={12},
  pages={123016},
  year={2009},
  publisher={APS}
}

@article{nc5chimento2010dbi,
  title={DBI models for the unification of dark matter and dark energy},
  author={Chimento, Luis P and Lazkoz, Ruth and Sendra, Irene},
  journal={General Relativity and Gravitation},
  volume={42},
  number={5},
  pages={1189--1209},
  year={2010},
  publisher={Springer}
}

@article{nc6cai2016dark,
  title={Dark matter superfluid and DBI dark energy},
  author={Cai, Rong-Gen and Wang, Shao-Jiang},
  journal={Physical Review D},
  volume={93},
  number={2},
  pages={023515},
  year={2016},
  publisher={APS}
}

@article{nc7chattopadhyay2010interaction,
  title={Interaction Between DBI-essence and other dark energies},
  author={Chattopadhyay, Surajit and Debnath, Ujjal},
  journal={International Journal of Theoretical Physics},
  volume={49},
  number={7},
  pages={1465--1480},
  year={2010},
  publisher={Springer}
}

@article{nc8calcagni2006tachyon,
  title={Tachyon dark energy models: Dynamics and constraints},
  author={Calcagni, Gianluca and Liddle, Andrew R},
  journal={Physical Review D—Particles, Fields, Gravitation, and Cosmology},
  volume={74},
  number={4},
  pages={043528},
  year={2006},
  publisher={APS}
}

@article{nc9bagla2003cosmology,
  title={Cosmology with tachyon field as dark energy},
  author={Bagla, JS and Jassal, Harvinder Kaur and Padmanabhan, T},
  journal={Physical Review D},
  volume={67},
  number={6},
  pages={063504},
  year={2003},
  publisher={APS}
}

@article{nc10copeland2005needed,
  title={What is needed of a tachyon if it is to be the dark energy?},
  author={Copeland, Edmund J and Garousi, Mohammad R and Sami, M and Tsujikawa, Shinji},
  journal={Physical Review D—Particles, Fields, Gravitation, and Cosmology},
  volume={71},
  number={4},
  pages={043003},
  year={2005},
  publisher={APS}
}

@article{nc11micheletti2010observational,
  title={Observational constraints on holographic tachyonic dark energy in interaction with dark matter},
  author={Micheletti, Sandro MR},
  journal={Journal of Cosmology and Astroparticle Physics},
  volume={2010},
  number={05},
  pages={009},
  year={2010},
  publisher={IOP Publishing}
}

@article{nc12sheykhi2012tachyon,
  title={Tachyon reconstruction of ghost dark energy},
  author={Sheykhi, A and Sadegh Movahed, M and Ebrahimi, E},
  journal={Astrophysics and Space Science},
  volume={339},
  number={1},
  pages={93--99},
  year={2012},
  publisher={Springer}
}

@article{Kehayias:2019gir,
    author = "Kehayias, John and Scherrer, Robert J.",
    title = "{New generic evolution for $k$ -essence dark energy with $w \approx -1$}",
    eprint = "1905.05628",
    archivePrefix = "arXiv",
    primaryClass = "gr-qc",
    doi = "10.1103/PhysRevD.100.023525",
    journal = "Phys. Rev. D",
    volume = "100",
    number = "2",
    pages = "023525",
    year = "2019"
}

@article{Chiba:2009nh,
    author = "Chiba, Takeshi and Dutta, Sourish and Scherrer, Robert J.",
    title = "{Slow-roll k-essence}",
    eprint = "0906.0628",
    archivePrefix = "arXiv",
    primaryClass = "astro-ph.CO",
    doi = "10.1103/PhysRevD.80.043517",
    journal = "Phys. Rev. D",
    volume = "80",
    pages = "043517",
    year = "2009"
}

@article{Das:2006cm,
    author = "Das, Rupam and Kephart, Thomas W. and Scherrer, Robert J.",
    title = "{Tracking quintessence and k-essence in a general cosmological background}",
    eprint = "gr-qc/0609014",
    archivePrefix = "arXiv",
    doi = "10.1103/PhysRevD.74.103515",
    journal = "Phys. Rev. D",
    volume = "74",
    pages = "103515",
    year = "2006"
}

@article{Hu1998,
  author = "Hu, Wayne",
  title = "{Structure Formation with Generalized Dark Matter}",
  journal = "Astrophys. J.",
  volume = "506",
  year = "1998",
  pages = "485",
  eprint = "astro-ph/9801234",
  archivePrefix = "arXiv"
}

@article{Planck2018, 
  title={Planck 2018 results. VI. Cosmological parameters},
  volume={641}, ISSN={0004-6361},
  DOI={10.1051/0004-6361/201833910},
  journal={Astronomy and Astrophysics}, publisher={EDP Sciences},
  author={Aghanim, N. and Ghosh, T. and Bock, J. J. and Crill, B. P. and Doré, O. and Hildebrandt, S. R. and Rocha, G. and Górski, K. M. and Lawrence, C. R. and Mitra, S. and Roudier, G. and Planck Collaboration},
  year={2020},
  month={Sep},
  pages={Art. No. A6} 
}

@onlie{Gurrola:2026fqx,
    author = "Gurrola, Alfredo and Scherrer, Robert J. and Trivedi, Oem",
    title = "{Probing the Sound Speed of Dark Energy with a Lunar Laser Interferometer}",
    eprint = "2601.22084",
    archivePrefix = "arXiv",
    primaryClass = "astro-ph.CO",
    month = "1",
    year = "2026"
}

@online{Gurrola:2026nks,
    author = "Gurrola, Alfredo and Scherrer, Robert J. and Trivedi, Oem",
    title = "{Probing Dark Energy on the Moon}",
    eprint = "2603.04841",
    archivePrefix = "arXiv",
    primaryClass = "astro-ph.CO",
    month = "3",
    year = "2026"
}

@article{Vasilev:2024deshiftsymm,
    author = "Vasilev, Teodor B. and Bouhmadi-L\'{o}pez, Mariam and Mart\'{i}n-Moruno, Prado",
    title = "{Dark energy with a shift-symmetric scalar field: Obstacles, loophole hunting and dead ends}",
    eprint = "2406.12576",
    archivePrefix = "arXiv",
    primaryClass = "gr-qc",
    doi = "10.1016/j.dark.2024.101679",
    journal = "Physics of the Dark Universe",
    volume = "46",
    pages = "101679",
    year = "2024"
}

@article{Deffayet:2010dekgb,
    author = "Deffayet, C\'{e}dric and Pujol\`{a}s, Oriol and Sawiki, Ignacy and Vikman, Alexander",
    title = "{Imperfect dark energy from kinetic gravity braiding}",
    eprint = "1008.0048",
    archivePrefix = "arXiv",
    primaryClass = "hep-th",
    doi = "10.1088/1475-7516/2010/10/026",
    journal = "Journal of Cosmology and Astroparticle Physics",
    volume = "2010",
    year = "2010"
}

@article{Finelli:2018shiftsymmcosm,
    author = "Finelli, Bernardo and Goon, Garrett and Pajer, Enrico and Santoni, Luca",
    title = "{The effective theory of shift-symmetric cosmologies}",
    eprint = "1802.01580",
    archivePrefix = "arXiv",
    primaryClass = "hep-th",
    doi = "10.1088/1475-7516/2018/05/060",
    journal = "Journal of Cosmology and Astroparticle Physics",
    volume = "2018",
    year = "2018"
}

@article{Chiang:supergravquint,
    author = "Chiang, Chien-I and Murayama, Hitoshi",
    title = "{Building Supergravity Quintessence Model}",
    journal = "arXiv preprint arXiv:1808.02279",
    year = "2018"
}

@article{Berbig:2025DESIquint,
    author = "Berbig, Maximilian",
    title = "{Kick it like DESI: PNGB quintessence with a dynamically generated initial velocity}",
    eprint = "2412.07418",
    archivePrefix = "arXiv",
    primaryClass = "astro-ph.CO",
    doi = "10.1088/1475-7516/2025/03/015",
    journal = "Journal of Cosmology and Astroparticle Physics",
    year = "2025"
}

@article{Kaneta:2024pngbinflation,
    author = "Kaneta, Kunio and Lee, Sung Mook and Oda, Kin-ya and Takahashi, Tomo",
    title = "{Pseudo-Nambu-Goldstone boson production from inflation coupling during reheating}",
    eprint = "2406.09045",
    archivePrefix = "arXiv",
    primaryClass = "astro-ph",
    doi = "10.1088/1475-7516/2024/11/058",
    journal = "Journal of Cosmology and Astroparticle Physics",
    volume = "2024",
    year = "2024"
}

@article{Sikivie:1983ip,
  author = {Sikivie, Pierre},
  title = {Experimental Tests of the Invisible Axion},
  journal = {Phys. Rev. Lett.},
  volume = {51},
  pages = {1415},
  year = {1983}
}

@article{Kim:1986ax,
  author = {Kim, Jihn E.},
  title = {Light pseudoscalars, particle physics and cosmology},
  journal = {Phys. Rept.},
  volume = {150},
  pages = {1-177},
  year = {1987}
}

@review{Marsh:2015xka,
  author = {Marsh, David J. E.},
  title = {Axion Cosmology},
  journal = {Phys. Rept.},
  volume = {643},
  pages = {1-79},
  year = {2016},
  eprint = {1510.07633},
  archivePrefix = {arXiv},
  primaryClass = {astro-ph.CO}
}

@book{Ziman:1960,
  author = {Ziman, J. M.},
  title = {Electrons and Phonons: The Theory of Transport Phenomena in Solids},
  publisher = {Oxford University Press},
  year = {1960}
}

@book{Kittel:2004,
  author = {Kittel, Charles},
  title = {Introduction to Solid State Physics},
  edition = {8th},
  publisher = {Wiley},
  year = {2004}
}

@article{Braaten:1990it,
  author = {Braaten, Eric and Pisarski, Robert D.},
  title = {Soft amplitudes in hot gauge theories: A general analysis},
  journal = {Nucl. Phys. B},
  volume = {337},
  pages = {569-634},
  year = {1990}
}

@book{Kapusta:2006pm,
  author = {Kapusta, Joseph I. and Gale, Charles},
  title = {Finite-Temperature Field Theory: Principles and Applications},
  publisher = {Cambridge University Press},
  edition = {2nd},
  year = {2006}
}

@book{Mukhanov:1990me,
  author = {Mukhanov, Viatcheslav F. and Feldman, H. A. and Brandenberger, Robert H.},
  title = {Theory of cosmological perturbations},
  journal = {Phys. Rept.},
  volume = {215},
  pages = {203-333},
  year = {1992}
}

@book{Weinberg:2008zzc,
  author = {Weinberg, Steven},
  title = {Cosmology},
  publisher = {Oxford University Press},
  year = {2008}
}

@article{Avsajanishvili:2024obsconst_dynamde,
    author = "Avsajanishvili, Olga and Chitov, Gennady Y. and Kahniashvili, Tina and Mandal, Sayan and Samushia, Lado",
    title = "{Observational Constraints on Dynamical Dark Energy Models}",
    eprint = "2310.16911",
    archivePrefix = "arXiv",
    primaryClass = "astro-ph.CO",
    doi = "10.3390/universe10030122",
    journal = "Universe",
    volume = "2024",
    year = "2024"
}

@article{Alwall:2014madgraph5,
    author = "J. Alwall and others",
    title = "{The automated computation of tree-level and next-to-leading order differential cross sections, and their matching to parton shower simulations}",
    eprint = "1405.0301",
    archivePrefix = "arXiv",
    primaryClass = "hep-ph",
    doi = "10.1007/JHEP07(2014)079",
    journal = "Journal of High Energy Physics",
    volume = "2014",
    year = "2014",
}

@article{Conte:2013madanalysis5,
    author = "Eric Conte and Benjamin Fuks and Guillaume Serret",
    title = "{\textsc{MadAnalysis} 5, a user-friendly framework for collider phenomenology}",
    eprint = "1206.1599",
    archivePrefix = "arXiv",
    primaryClass = "hep-ph",
    doi = "10.1016/j.cpc.2012.09.009",
    journal = "Computer Physics Communications",
    volume = "184",
    year = "2013",
}

@article{CMS:2024zqs,
    author = "Hayrapetyan, Aram and others",
    collaboration = "CMS",
    title = "{Dark sector searches with the CMS experiment}",
    eprint = "2405.13778",
    archivePrefix = "arXiv",
    primaryClass = "hep-ex",
    reportNumber = "CMS-EXO-23-005, CERN-EP-2024-106",
    doi = "10.1016/j.physrep.2024.09.013",
    journal = "Phys. Rept.",
    volume = "1115",
    pages = "448--569",
    year = "2025"
}

@article{Babichev:2018twg,
    author = "Babichev, Eugeny and Ramazanov, Sabir and Vikman, Alexander",
    title = "{Recovering $P(X)$ from a canonical complex field}",
    eprint = "1807.10281",
    archivePrefix = "arXiv",
    primaryClass = "gr-qc",
    reportNumber = "LPT-Orsay-18-89",
    doi = "10.1088/1475-7516/2018/11/023",
    journal = "JCAP",
    volume = "11",
    pages = "023",
    year = "2018"
}

@article{Glavan:2025khe,
    author = "Glavan, Dra{\v{z}}en and Vikman, Alexander and Zlosnik, Tom",
    title = "{On the Bondi accretion of a self-interacting complex scalar field}",
    journal = "arXiv preprint arXiv:2511.04650",
    eprint = "2511.04650",
    archivePrefix = "arXiv",
    primaryClass = "gr-qc",
    month = "11",
    year = "2025"
}

@article{Babichev:2007dw,
    author = "Babichev, Eugeny and Mukhanov, Viatcheslav and Vikman, Alexander",
    title = "{k-Essence, superluminal propagation, causality and emergent geometry}",
    eprint = "0708.0561",
    archivePrefix = "arXiv",
    primaryClass = "hep-th",
    reportNumber = "LMU-ASC-54-07",
    doi = "10.1088/1126-6708/2008/02/101",
    journal = "JHEP",
    volume = "02",
    pages = "101",
    year = "2008"
}

@article{OShea:2024jjw,
    author = "O'Shea, Tom{\'a}s and Davis, Anne-Christine and Giannotti, Maurizio and Vagnozzi, Sunny and Visinelli, Luca and Vogel, Julia K.",
    title = "{Solar chameleons: Novel channels}",
    eprint = "2406.01691",
    archivePrefix = "arXiv",
    primaryClass = "hep-ph",
    doi = "10.1103/PhysRevD.110.063027",
    journal = "Phys. Rev. D",
    volume = "110",
    number = "6",
    pages = "063027",
    year = "2024"
}

@article{Yuan:2025twx,
    author = "Yuan, Guan-Wen and Davis, Anne-Christine and Giannotti, Maurizio and Vagnozzi, Sunny and Visinelli, Luca and Vogel, Julia K.",
    title = "{Direct detection of solar chameleons with electron recoil data from XENONnT}",
    eprint = "2511.01655",
    archivePrefix = "arXiv",
    primaryClass = "hep-ph",
    doi = "10.1103/8mpf-5s3k",
    journal = "Phys. Rev. D",
    volume = "113",
    number = "12",
    pages = "123024",
    year = "2026"
}

@article{Vagnozzi:2019kvw,
    author = "Vagnozzi, Sunny and Visinelli, Luca and Mena, Olga and Mota, David F.",
    title = "{Do we have any hope of detecting scattering between dark energy and baryons through cosmology?}",
    eprint = "1911.12374",
    archivePrefix = "arXiv",
    primaryClass = "gr-qc",
    doi = "10.1093/mnras/staa311",
    journal = "Mon. Not. Roy. Astron. Soc.",
    volume = "493",
    number = "1",
    pages = "1139--1152",
    year = "2020"
}

@article{Ferlito:2022mok,
    author = "Ferlito, Fulvio and Vagnozzi, Sunny and Mota, David F. and Baldi, Marco",
    title = "{Cosmological direct detection of dark energy: Non-linear structure formation signatures of dark energy scattering with visible matter}",
    eprint = "2201.04528",
    archivePrefix = "arXiv",
    primaryClass = "astro-ph.CO",
    doi = "10.1093/mnras/stac649",
    journal = "Mon. Not. Roy. Astron. Soc.",
    volume = "512",
    number = "2",
    pages = "1885--1905",
    year = "2022"
}

@article{Bamba:2012cp,
    author = "Bamba, Kazuharu and Capozziello, Salvatore and Nojiri, Shin'ichi and Odintsov, Sergei D.",
    title = "{Dark energy cosmology: the equivalent description via different theoretical models and cosmography tests}",
    eprint = "1205.3421",
    archivePrefix = "arXiv",
    primaryClass = "gr-qc",
    doi = "10.1007/s10509-012-1181-8",
    journal = "Astrophys. Space Sci.",
    volume = "342",
    pages = "155--228",
    year = "2012"
}

@article{Colgain:2025nzf,
    author = "Colg{\'a}in, Eoin {\'O}. and Pourojaghi, Saeed and Sheikh-Jabbari, M. M. and Yin, Lu",
    title = "{How much has DESI dark energy evolved since DR1?}",
    eprint = "2504.04417",
    archivePrefix = "arXiv",
    primaryClass = "astro-ph.CO",
    doi = "10.1016/j.dark.2026.102268",
    journal = "Phys. Dark Univ.",
    volume = "52",
    pages = "102268",
    year = "2026"
}

@article{Sawicki:2024ryt,
    author = "Sawicki, Ignacy and Trenkler, Georg and Vikman, Alexander",
    title = "{Causality and stability from acoustic geometry}",
    eprint = "2412.21169",
    archivePrefix = "arXiv",
    primaryClass = "gr-qc",
    doi = "10.1007/JHEP10(2025)227",
    journal = "JHEP",
    volume = "10",
    pages = "227",
    year = "2025"
}

@article{DESI:2025zgx,
    author = "Abdul Karim, M. and others",
    collaboration = "DESI",
    title = "{DESI DR2 results. II. Measurements of baryon acoustic oscillations and cosmological constraints}",
    eprint = "2503.14738",
    archivePrefix = "arXiv",
    primaryClass = "astro-ph.CO",
    reportNumber = "FERMILAB-PUB-25-0169-PPD",
    doi = "10.1103/tr6y-kpc6",
    journal = "Phys. Rev. D",
    volume = "112",
    number = "8",
    pages = "083515",
    year = "2025"
}

@article{Scherer:2025esj,
    author = "Scherer, Mateus and Sabogal, Miguel A. and Nunes, Rafael C. and De Felice, Antonio",
    title = "{Challenging the {\ensuremath{\Lambda}}CDM model: 5{\ensuremath{\sigma}} evidence for a dynamical dark energy late-time transition}",
    eprint = "2504.20664",
    archivePrefix = "arXiv",
    primaryClass = "astro-ph.CO",
    doi = "10.1103/n86r-sjgm",
    journal = "Phys. Rev. D",
    volume = "112",
    number = "4",
    pages = "043513",
    year = "2025"
}

@article{Silva:2025twg,
    author = "Silva, Emanuelly and Nunes, Rafael C.",
    title = "{Testing signatures of phantom crossing through full-shape galaxy clustering analysis}",
    eprint = "2507.13989",
    archivePrefix = "arXiv",
    primaryClass = "astro-ph.CO",
    doi = "10.1088/1475-7516/2025/11/078",
    journal = "JCAP",
    volume = "11",
    pages = "078",
    year = "2025"
}

@article{Silva:2025hxw,
    author = "Silva, Emanuelly and Sabogal, Miguel A. and Scherer, Mateus and Nunes, Rafael C. and Di Valentino, Eleonora and Kumar, Suresh",
    title = "{New constraints on interacting dark energy from DESI DR2 BAO observations}",
    eprint = "2503.23225",
    archivePrefix = "arXiv",
    primaryClass = "astro-ph.CO",
    doi = "10.1103/qqc6-76z4",
    journal = "Phys. Rev. D",
    volume = "111",
    number = "12",
    pages = "123511",
    year = "2025"
}

@article{Shah:2025ayl,
    author = "Shah, Rahul and Mukherjee, Purba and Pal, Supratik",
    title = "{Interacting dark sectors in light of DESI DR2}",
    eprint = "2503.21652",
    archivePrefix = "arXiv",
    primaryClass = "astro-ph.CO",
    doi = "10.1093/mnras/staf1442",
    journal = "Mon. Not. Roy. Astron. Soc.",
    volume = "542",
    number = "4",
    pages = "2936--2942",
    year = "2025"
}

@article{Pan:2025qwy,
    author = "Pan, Supriya and Paul, Sivasish and Saridakis, Emmanuel N. and Yang, Weiqiang",
    title = "{Interacting dark energy after DESI DR2: A challenge for the {\ensuremath{\Lambda}}CDM paradigm?}",
    eprint = "2504.00994",
    archivePrefix = "arXiv",
    primaryClass = "astro-ph.CO",
    doi = "10.1103/5y21-k39n",
    journal = "Phys. Rev. D",
    volume = "113",
    number = "2",
    pages = "023515",
    year = "2026"
}

@article{Argyropoulos:2024yxo,
    author = "Argyropoulos, Spyros and Haisch, Ulrich and Kalaitzidou, Ilia",
    title = "{Novel collider signatures in the type-I 2HDM+a model}",
    eprint = "2404.05704",
    archivePrefix = "arXiv",
    primaryClass = "hep-ph",
    reportNumber = "MPP-2024-66",
    doi = "10.1007/JHEP07(2024)263",
    journal = "JHEP",
    volume = "07",
    pages = "263",
    year = "2024"
}

@article{Scherrer:2004kinetickess,
    author = "Scherrer, Robert J.",
    title = "{Purely kinetic k-essence as unified dark matter}",
    eprint = "astro-ph/0402316",
    archivePrefix = "arXiv",
    primaryClass = "astro-ph",
    doi = "10.1103/PhysRevLett.93.011301",
    journal = "Phys. Rev. Lett.",
    volume = "93",
    pages = "011301",
    year = "2004"
}

@article{Cordero:2016kink-esstop,
    author = "Cordero, Ruben and Gonzalez, Eduardo L. and Queijeiro, Alfonso",
    title = "{An equation of state for purely kinetic k-essence inspired by cosmic topological defects}",
    eprint = "1608.06540",
    archivePrefix = "arXiv",
    primaryClass = "gr-qc",
    doi = "10.1140/epjc/s10052-017-4913-7",
    journal = "Eur. Phys. J. C",
    volume = "77",
    number = "413",
    year = "2016"
}

@article{PDG:2026review,
    author = {F. Takahashi and others},
    collaboration = {Particle Data Group},
    title = {Review of Particle Physics},
    journal = {International Journal of Modern Physics A},
    volume = {41},
    pages = {2630011},
    year = {2026}
}
\end{document}